\documentclass[aps,prl,showpacs,twocolumn,superscriptaddress,floatfix]{revtex4-2}

\usepackage{times,xspace}
\usepackage{graphicx,graphics,color,epsfig}
\usepackage{amsmath}
\usepackage{amssymb}
\usepackage{amsfonts}
\usepackage{amsfonts}
\usepackage{epstopdf}
\usepackage{bm}
\usepackage{hyperref}
\usepackage{enumitem}
\usepackage{color}
\usepackage[normalem]{ulem}
\usepackage{xcolor}

\def\be{\begin{equation}}
\def\ee{\end{equation}}
\def\ba{\begin{eqnarray}}
\def\ea{\end{eqnarray}}

\newcommand{\sur}[1]{{\color{orange!75!black}{#1}}}
\newcommand{\qs}[1]{{\color{magenta}{#1}}}

\begin{document}

\title{Nonperturbative Nonlinear Hall Effect in Nonequilibrium Steady States}

\author{Lei Geng}
\email{lei.geng@unifr.ch}
\affiliation{Department of Physics, University of Fribourg, Fribourg-1700, Switzerland}
\author{Martin Eckstein}
\affiliation{Institute of Theoretical Physics, University of Hamburg, 20355 Hamburg, Germany}
\affiliation{The Hamburg Centre for Ultrafast Imaging, Hamburg, Germany}
\author{Lei Chen}
\affiliation{Department of Physics and Astronomy, Extreme Quantum Materials Alliance, Smalley-Curl Institute, 
Rice University, Houston, Texas 77005, USA}
\affiliation{Department of Physics and Astronomy, Stony Brook University, Stony Brook, New York 11794, USA}
\author{Shouvik Sur}
\affiliation{Department of Physics and Astronomy, Extreme Quantum Materials Alliance, Smalley-Curl Institute, 
Rice University, Houston, Texas 77005, USA}
\author{Silke Paschen}
\affiliation{Institute of Solid State Physics, TU Wien, 1040 Vienna, Austria}
\author{Qimiao Si}
\affiliation{Department of Physics and Astronomy, Extreme Quantum Materials Alliance, Smalley-Curl Institute, 
Rice University, Houston, Texas 77005, USA}
\author{Philipp Werner}
\email{philipp.werner@unifr.ch}
\affiliation{Department of Physics, University of Fribourg, Fribourg-1700, Switzerland}

\date{\today}
 
\begin{abstract}
The nonlinear Hall effect  in quantum materials has attracted broad interest, yet most existing studies focus on the weak-field, perturbative regime.
Here we develop a nonperturbative approach based on nonequilibrium steady-state Green's functions for dc-field-driven  lattice systems, with dissipation and interactions incorporated through self-energies beyond the constant relaxation-time approximation and interband transitions treated alongside their intraband counterparts.
Applied to a two-band semimetal model, our approach provides direct access to the strong-field Hall response beyond the nonperturbative crossover where the edge of the nonequilibrium distribution reaches Berry-curvature hot spots, a regime in which constant relaxation-time estimates and Berry curvature dipole calculations become unreliable.
We further demonstrate that interaction and electron-phonon self-energies within dynamical mean-field theory can substantially change the Hall signal.
Our framework enables quantitative simulations of nonequilibrium nonlinear Hall phenomena and provides guidance for strong-field transport experiments.
\end{abstract}

\maketitle

The interplay between topology, nonequilibrium and correlations is an outstanding challenge in quantum condensed matter physics. For example, electrical transport can reveal signatures of topology, altough in correlated settings, theoretical studies are generally difficult.  
Consider the Hall effect, which is a central paradigm of condensed matter physics, linking transverse transport to the symmetry, topology, and geometry of electronic states.
In the linear regime, a Hall current without an external magnetic field requires broken time-reversal symmetry, as in the anomalous Hall effect.
This constraint is lifted beyond linear response: in time-reversal-invariant but inversion-breaking systems, a longitudinal electric field can drive a transverse current at second order, giving rise to the nonlinear Hall effect~\cite{Sodemann2015,Du2021Review,bandyopadhyay2024non}.
The simplest picture is the Berry curvature dipole—the electric field shifts the distribution of occupied states, producing a net anomalous velocity even though the Berry curvature integrates to zero in equilibrium~\cite{Sodemann2015,deyo2009,moore2010}.
This mechanism yields rectified and second-harmonic Hall responses, and it can be used to detect momentum-space Berry curvature in nonmagnetic materials.

The nonlinear Hall effect has rapidly evolved from a theoretical proposal into an experimental probe of quantum materials.
Second-order Hall signals were observed in inversion-breaking, time-reversal-symmetric systems,
including bilayer and few-layer WTe$_2$~\cite{Ma2019Nature,Kang2019NatMater},
the ultrathin Weyl semimetal TaIrTe$_4$~\cite{Kumar2021NatNano}, and the 3D Weyl-Kondo
semimetal Ce$_3$Bi$_4$Pd$_3$~\cite{dzsaber2021giant}.
The observation of a giant nonlinear Hall response in Ce$_3$Bi$_4$Pd$_3$ marked a key advance,
providing compelling bulk transport evidence for the Weyl-Kondo semimetal state
~\cite{lai2018weyl,dzsaber2017kondo,Kofuji2021}.
Higher-order responses have further been analyzed in terms of Berry-curvature multipoles and
related geometric quantities~\cite{Zhang2023PRB,Wang2022NSR,Lai2021NatNano}.

Beyond the Berry-curvature-dipole picture, quantum-kinetic and
diagrammatic approaches have shown that disorder-induced side-jump and
skew-scattering processes contribute at the same order and can
substantially modify the weak-field nonlinear Hall
response~\cite{Du2019NatCommun,Nandy2019PRB,Xiao2019PRB,Du2021NatCommun}.
Perturbative quantum formulations have further clarified their
connection to semiclassical Boltzmann theory and the roles of
dissipation, interband coherence, and multiband
geometry~\cite{Kaplan2023SciPost,Michishita2021,Kofuji2021,
Michishita2022PRB,Tokura2018NatCommun,fang2025nonlinear,Wacker1999}.
Relaxation-time-independent intrinsic mechanisms have also been
formulated in terms of Berry-connection polarizability and related
geometric quantities, particularly in magnetic
systems~\cite{Gao2014PRL,Wang2021PRL,Liu2021PRL,
Kaplan2023NatCommun,Kaplan2024PRL,Gao2023Science,Wang2023Nature}.

While these developments reveal a rich interplay of Berry curvature, quantum metric, disorder, and dissipation in perturbative nonlinear Hall transport, the nonperturbative strong-field regime remains underexplored. 
First experimental evidence for this strong-field regime was found in Ce$_3$Bi$_4$Pd$_3$, where the Hall resistivity strongly deviates from Berry-curvature-dipole scaling~\cite{dzsaber2021giant}. Similar behavior was also found in Boltzmann-equation studies,  when the driven distribution crosses Berry-curvature hot spots~\cite{Sur2024FNE}. 
To systematically explore the nonperturbative regime, including the role of dissipation and electron correlations,
we develop a steady-state framework for dc-field-driven lattice systems, based on nonequilibrium Green's functions (NEGF).
While NEGF calculations have recently been applied to ultrafast nonlinear Hall dynamics~\cite{dendzik2026ultrafast}, we focus on time-translationally invariant steady states under a constant electric field, following approaches developed for photodoped~\cite{Li2021,Yuta2025,geng2026high,geng2026photoinduced,geng2025third} and electric-field-driven~\cite{Yuta2018,Li2015PRL,Han2018PRB,Aron2012,Aron2012b,Aron2013} systems. 
In this framework, the uniform dc field is incorporated through the gauge-covariant Wigner representation, where the electric field enters via the Moyal star product~\cite{groenewold1946principles,moyal1949quantum}.
The natural momentum variable in this representation is the gauge-covariant momentum $\boldsymbol{\kappa}$, which reduces to the usual momentum in the absence of the field \footnote{
The dc field is incorporated through the gauge-covariant Wigner representation, with \(\boldsymbol{\kappa}\) denoting the gauge-covariant momentum.
}.
Dissipation is introduced through the coupling to external baths, which stabilizes a nonequilibrium steady state. In contrast to semiclassical Boltzmann-equation approaches, the NEGF formulation retains interband coherence, treats the driving field nonperturbatively, and incorporates 
interactions
and electron-phonon coupling through frequency-dependent self-energies.
It does not rely on a quasi-particle approximation 
and allows to compute the Hall current without relying on the Berry-curvature-dipole picture.
The full details of the derivation of the formalism and the numerical implementation are provided in the Supplemental Material~\cite{SM}.

\begin{figure}[tbp]
    \centering
    \includegraphics[width=\columnwidth]{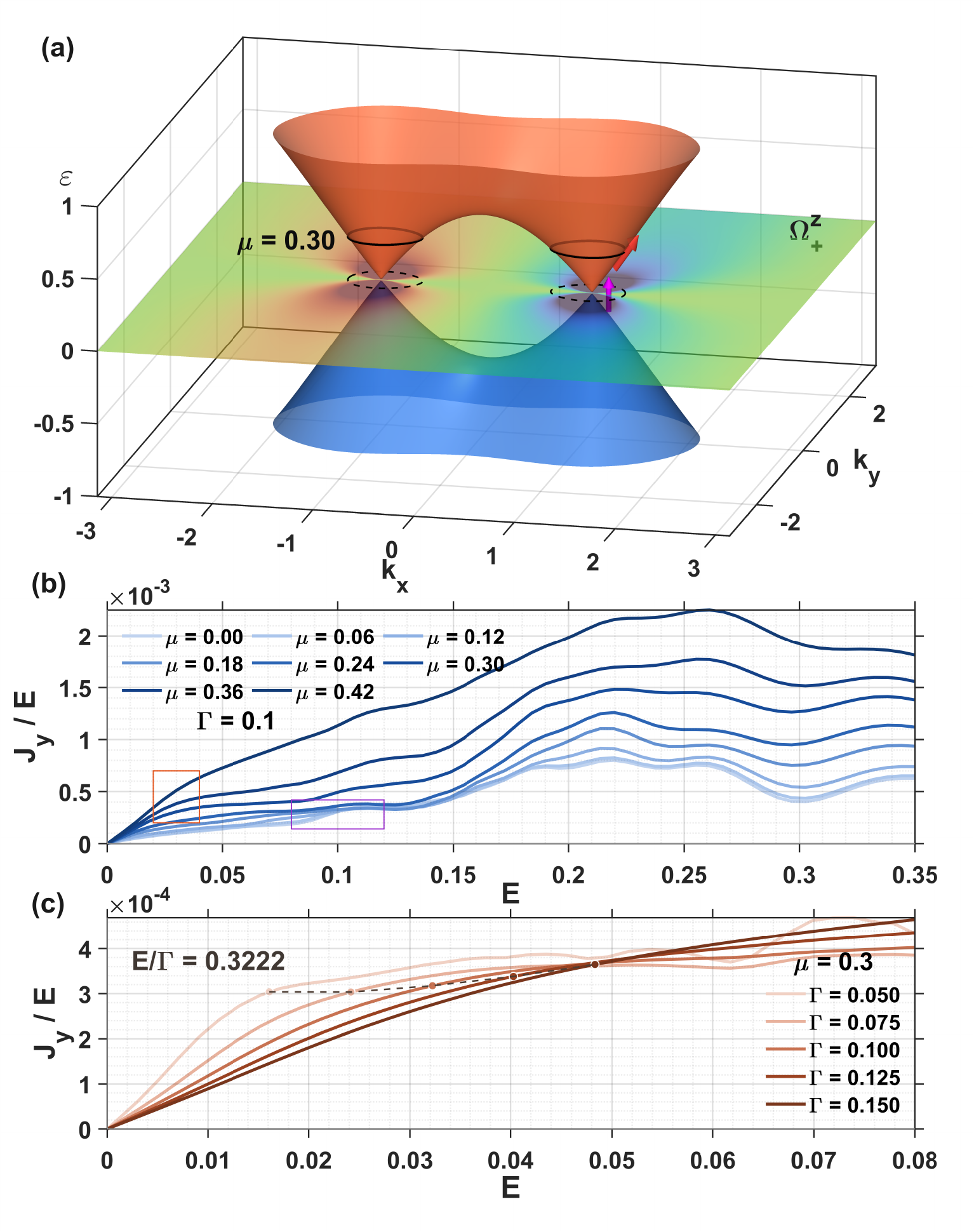}
\caption{
(a) Band structure of the two-band model in the $(k_x,k_y)$ plane, with the Berry curvature of the upper band projected onto the $\varepsilon=0$ plane and the Fermi contour for $\mu=0.30$ indicated. The red and magenta arrow denote representative intraband and interband electron transfer processes near the right  cone under a constant electric field. 
(b) Field-normalized transverse current $J_y/E$ as a function of the driving field $E$ for several chemical potentials $\mu$ at fixed reservoir damping parameter $\Gamma=0.1$. 
(c) $J_y/E$ as a function of $E$ for several reservoir damping parameters $\Gamma$ at fixed chemical potential $\mu=0.30$. The dots connected by the dashed line indicate the condition $E/\Gamma=\Delta k_x$, where $\Delta k_x=0.3222$ is the distance between the nodal point and the equilibrium Fermi surface.
}
\label{fig_1}
\end{figure}

We apply this framework to a minimal two-band lattice model that contains the key ingredients for the nonlinear Hall effect in a time-reversal-invariant but inversion-broken system.
The Hamiltonian is 
\begin{equation}
H(\mathbf{k})
=
(\cos k_x-D_x)\sigma_x
+
\sin k_y\,\sigma_y
+ (\cos k_y-D_y)\sigma_z ,
\label{eq:model}
\end{equation}
where $\sigma_{\alpha=x,y,z}$ are Pauli matrices in orbital space \footnote{
In a minimal linearized $k\cdot p$ model of a single cone, a tilt term proportional to $\sigma_0$ is needed to reproduce the asymmetric nonequilibrium distribution near the cone. In Eq.~\eqref{eq:model}, this effect arises from the full lattice momentum dependence, and no explicit $\sigma_0$ term is required.
}.
In the following, we use $D_x=0.5$ and $D_y=1.0$, and set the lattice constant and hopping to unity.
For these parameters, the model describes a semimetal with two
nodal points. This minimal two-dimensional model is designed to mimic
the physics relevant to three-dimensional Weyl semimetals.
In particular, each nodal point is a twofold band crossing associated
with a pronounced Berry-curvature distribution and in this sense is
Weyl-like. The band dispersion is shown in Fig.~\ref{fig_1}(a),
together with the Berry-curvature distribution plotted on the
$\varepsilon=0$ plane. Because time-reversal symmetry is preserved,
the Berry curvature is odd in momentum and integrates to zero over
the Brillouin zone. It is strongly concentrated near the two nodal
points, where it has opposite signs.

We consider the response to a dc electric field applied along the $x$ direction. In the semiclassical picture, the Berry curvature contributes to the transverse current through the anomalous velocity,
$\mathbf{v}_{\rm anom}=-\mathbf{E}\times\boldsymbol{\Omega}_n(\boldsymbol{k})$.
Thus, the Hall response is obtained from a Berry-curvature-weighted occupation~\cite{gradhand2012first},
\begin{equation}
\frac{J_y}{E_x}
=
\sum_\nu
\int_{\rm BZ}\frac{d^2 k}{(2\pi)^2}
\, n_\nu(\boldsymbol{k})\Omega^z_\nu(\boldsymbol{k}) ,
\label{eq:anomalous_velocity}
\end{equation}
where $n_\nu(\boldsymbol{k})$ and $\Omega^z_\nu(\boldsymbol{k})$ are the occupation and Berry curvature of band $\nu$.
In equilibrium this integral vanishes in time-reversal-invariant systems, but an electric field can distort the occupation and make it finite.
In the weak-field limit, this becomes the Berry-curvature-dipole mechanism~\cite{Sodemann2015}, which is directly implemented in Boltzmann-equation treatments~\cite{Sur2024FNE,gradhand2012first}.
The NEGF formalism, by contrast, can evaluate the Hall current more generally using the velocity operator $v_y(\boldsymbol{\kappa})=\partial H(\boldsymbol{\kappa})/\partial \kappa_y$,
\begin{equation}
J_y
=
\int_{\rm BZ}\frac{d^2\kappa}{(2\pi)^2}
\,\mathrm{Tr}\!\left[
v_y(\boldsymbol{\kappa})\,\widetilde n(\boldsymbol{\kappa})
\right],
\label{eq:NEGF_jy}
\end{equation}
where $\widetilde n(\boldsymbol{\kappa})=-i\widetilde G^<(\tau=0,\boldsymbol{\kappa})=-i G^<(\tau=0,\boldsymbol{\kappa}+q\boldsymbol{A})$ is the gauge-covariant equal-time density matrix in the field-driven steady state, with \(G^<\) the lesser Green's function and \(\boldsymbol A\) the vector potential.

In Dirac or Weyl semimetals, the Fermi surface can lie close to a
nodal point, so that even a modest field can drive carriers into the
nodal region and make multiband effects important. The resulting
interband charge transfer, indicated by the magenta arrow in
Fig.~\ref{fig_1}(a), is not captured by the Berry-curvature-dipole
description, and Eq.~\eqref{eq:anomalous_velocity} no longer reproduces
the full Hall current~\cite{Culcer2017}. Although these interband
contributions can be included in a full second-order Kubo calculation
in the perturbative limit~\cite{Michishita2021,Michishita2022PRB}, at
$\mu=0$ and low temperature its integrand becomes sharply singular near
the nodal points, leading to severe momentum-grid convergence problems
in the numerical implementation. The NEGF formulation retains the full
multiband density matrix, exhibits much better momentum-grid convergence
for the present model, and yields the current response without
expanding in powers of the electric field.

To quantify the field and doping dependence, Fig.~\ref{fig_1}(b) shows $J_y/E$ for several chemical potentials.
The system is coupled to a zero-temperature orbital-independent
wide-band fermionic reservoir with chemical potential $\mu$ and
energy-independent retarded self-energy
$\Sigma_{\rm bath}^R=-i\Gamma$, with $\Gamma=0.1$. We refer to this
setup as the constant-$\Gamma$ reservoir model. The bath exchanges both
particles and energy with the system and provides a featureless
relaxation channel for stabilizing the driven steady state~\cite{Li2015PRL,Han2013}.
For the chosen $\Gamma$ parameter, $J_y/E$ is approximately linear in $E$ over
an appropriately defined low-field window, consistent with the
perturbative scaling $J_y\propto E^2$.
This behavior is also approximately observed at $\mu=0$, where no conventional Fermi surface exists and the response is instead driven by field-induced interband charge transfer near the cones. Results for smaller $\Gamma$ are discussed in the Supplemental Material.
As $\mu$ is increased, the magnitude of the Hall response grows.
For low and intermediate chemical potentials, the field dependence, especially at stronger fields, remains similar to the semimetallic case, indicating that interband processes still play an important role.

For larger chemical potentials, the curves bend near $E\simeq 0.03$, as highlighted by the orange box in Fig.~\ref{fig_1}(b).
This marks the onset of the nonperturbative drift mechanism: the field-driven Fermi-contour edge reaches the Berry-curvature hot spot near a 
nodal point~\cite{Sur2024FNE}.
Beyond this point, the weak-field Berry-curvature-dipole expansion is no longer controlled for the reasons discussed above, even if $J_y/E$ remains approximately linear over some field range.
At still stronger fields, especially when $E$ becomes comparable to the damping scale $\Gamma$, additional deviations appear, including the bump structure marked by the purple box in Fig.~\ref{fig_1}(b).
The response eventually enters a fully nonperturbative regime, where the Hall signal can even decrease with increasing field.
We attribute the small oscillatory features in $J_y/E$ to
Landau--Zener--St\"uckelberg interference between coherent upper- and
lower-band amplitudes~\cite{SHEVCHENKO20101,Lim2015}. A detailed
analysis is presented in the Supplemental Material~\cite{SM}.

We next examine the nonperturbative drift mechanism highlighted by the orange box.
We fix $\mu=0.3$ and compute $J_y/E$ for several damping rates $\Gamma$, as shown in Fig.~\ref{fig_1}(c).
For this chemical potential, the Fermi-contour edge is separated from the nearest nodal point by $\Delta \kappa_x=0.3222$.
Using the drift estimate $\Delta \kappa_x\simeq E/\Gamma$, the crossover is expected at $E^\ast\simeq \Gamma\Delta \kappa_x$.
The corresponding values are marked in Fig.~\ref{fig_1}(c) and connected by a dashed line.
Their agreement with the bending points of the Hall response supports the interpretation that the crossover is controlled by the field-driven shift of the Fermi contour to a Berry-curvature hot spot.

\begin{figure}[tbp]
    \centering
    \includegraphics[width=\columnwidth]{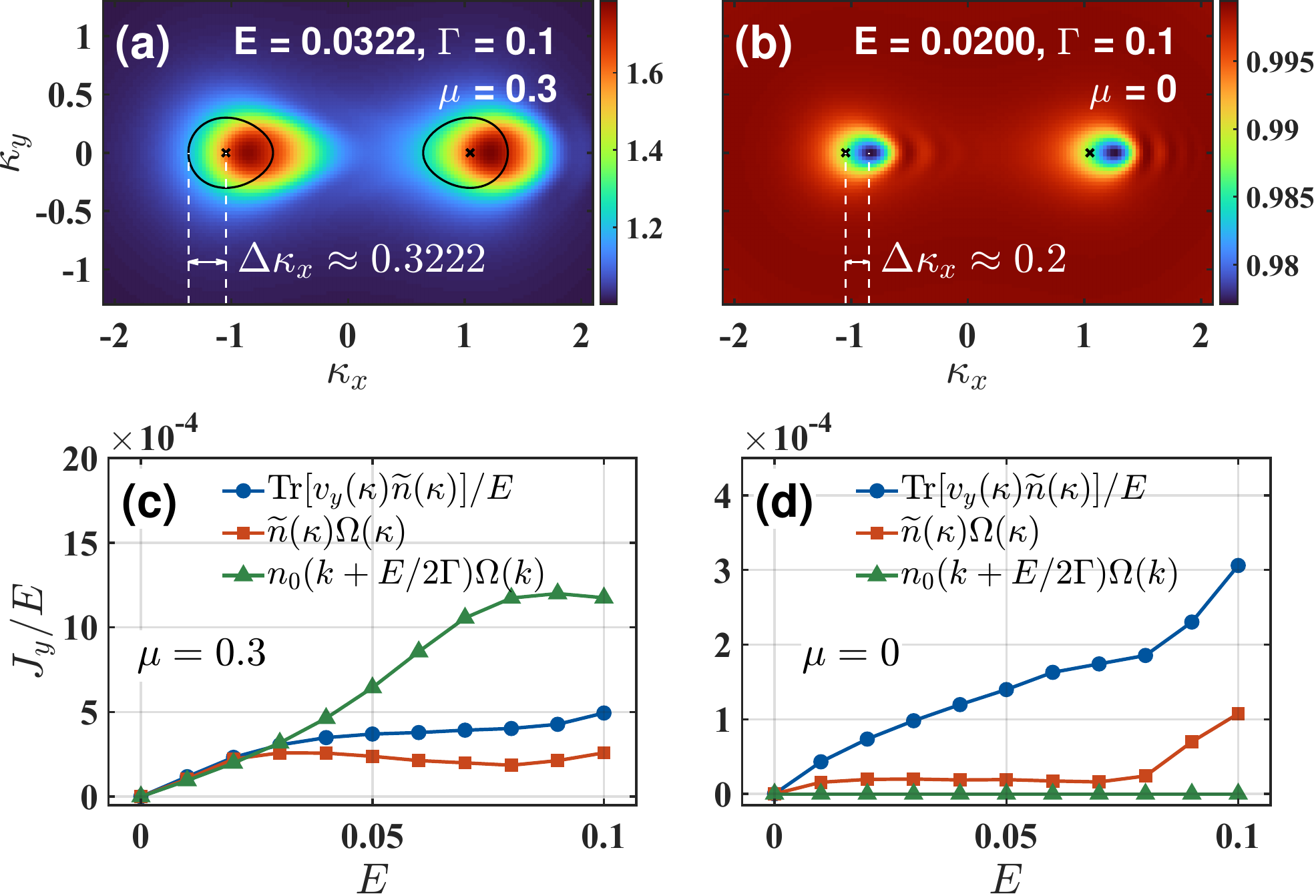}
\caption{
(a,b) Momentum-resolved occupation $\mathrm{Tr}\,\widetilde n(\boldsymbol{\kappa})$ for $\mu=0.3$ and $\mu=0$, respectively, at the indicated value of $E$. Black crosses mark the nodal points, and the black contour in (a) shows the Fermi surface. The dashed white lines indicate the characteristic momentum interval $\Delta \kappa_x$. 
(c,d) Hall-current response $J_y/E$ as a function of the electric field $E$ for $\mu=0.3$ and $\mu=0$, comparing the NEGF result $\mathrm{Tr}[v_y(\boldsymbol{\kappa})\widetilde n(\boldsymbol{\kappa})]/E$, the Berry-curvature estimate $\widetilde n(\boldsymbol{\kappa})\Omega(\boldsymbol{\kappa})$, and the shifted-equilibrium estimate $n_0(\mathbf{k}+E/2\Gamma)\Omega(\mathbf{k})$. All the calculations use $\Gamma=0.1$.
}
\label{fig_2}
\end{figure}

To visualize the momentum-space evolution associated with intra- and interband processes, we plot representative occupations $\mathrm{Tr}\,\widetilde n(\boldsymbol{\kappa})$ in Fig.~\ref{fig_2} for $\Gamma=0.1$.
We first consider the case with $\mu=0.3$, where the nonperturbative crossover is driven by intraband drift.
As shown in Fig.~\ref{fig_2}(a), at $E^\ast\simeq \Gamma\Delta \kappa_x=0.0322$, the occupation edge is displaced to the vicinity of the
nodal point.
This directly illustrates the drift mechanism responsible for the nonperturbative feature marked by the orange box in Fig.~\ref{fig_1}(b), in qualitative agreement with the picture proposed in Ref.~\cite{Sur2024FNE}.

We then turn to the semimetallic case $\mu=0$, where the response is driven by field-induced interband charge transfer.
Figure~\ref{fig_2}(b) shows two hole-like structures displaced from the corresponding nodal points by approximately $E/\Gamma$ along the field direction, accompanied by oscillatory tails \footnote{
The slight downward shift of the average occupation from $1.0$ originates from finite time- and momentum-grid discretization errors.
This nearly uniform offset does not affect the Hall current, which is computed directly from the converged density matrix.
}. 
Since vertical interband excitation at fixed momentum conserves $\mathrm{Tr}\,\widetilde n(\boldsymbol{\kappa})$, the observed hole-like structures reflect field- and dissipation-induced momentum-space redistribution, which can generate a Hall response in the anomalous-velocity picture through the opposite Berry curvatures of the two bands.

However, this does not imply that Eq.~\eqref{eq:anomalous_velocity} is quantitatively reliable for the Hall response induced by interband processes.
In the direct NEGF calculation, the Hall current is evaluated by Eq.~\eqref{eq:NEGF_jy}. For comparison, we transform the density matrix to the field-free band basis and evaluate Eq.~\eqref{eq:anomalous_velocity}.
The results for different $\mu$ are shown in Figs.~\ref{fig_2}(c) and \ref{fig_2}(d).
At nonzero chemical potential and weak fields, the Berry-curvature-weighted estimate agrees well with the full NEGF current.
Once interband processes become important, however, the discrepancy becomes pronounced.
This is the case both for $\mu=0$, where the response is intrinsically interband, and for the nonperturbative regime at $\mu=0.3$, where the Fermi contour approaches the 
nodal point and interband effects become significant.
Thus, even when the same density matrix is used, Eq.~\eqref{eq:anomalous_velocity} fails to reproduce the full Hall current when interband processes play a role. This comparison also highlights a limitation of the Boltzmann-equation method based on the Berry curvature dipole.

We next compare the NEGF results with a semiclassical
constant-relaxation-time estimate. In the weak-field limit, the average
distribution shift produced by the constant-$\Gamma$ reservoir model can be
approximately matched by choosing $\tau_B=1/(2\Gamma)$, corresponding
to a rigid displacement $\Delta\kappa=E/(2\Gamma)$~\cite{SM}. Using
this identification, Figs.~\ref{fig_2}(c) and \ref{fig_2}(d) compare
the NEGF current with a Berry-curvature estimate obtained from the
rigidly shifted equilibrium density matrix. This construction captures
only intraband redistribution and therefore gives no Hall response at
$\mu=0$, where the response is driven by interband processes.

For $\mu=0.3$, the shifted-distribution estimate agrees reasonably
well with the NEGF result at weak fields but predicts a different
crossover field. In the constant-$\Gamma$ reservoir model, the relevant
displacement of states at the Fermi-contour edge is $E/\Gamma$, giving
$E_\Gamma\simeq\Gamma\Delta\kappa_x$, whereas the matched
constant-relaxation-time model gives
$E_{\tau_B}\simeq2\Gamma\Delta\kappa_x$. Thus, the hot-spot criterion is
independent of the dissipation model, but its quantitative crossover
field is not.

\begin{figure}[tbp]
    \centering
    \includegraphics[width=\columnwidth]{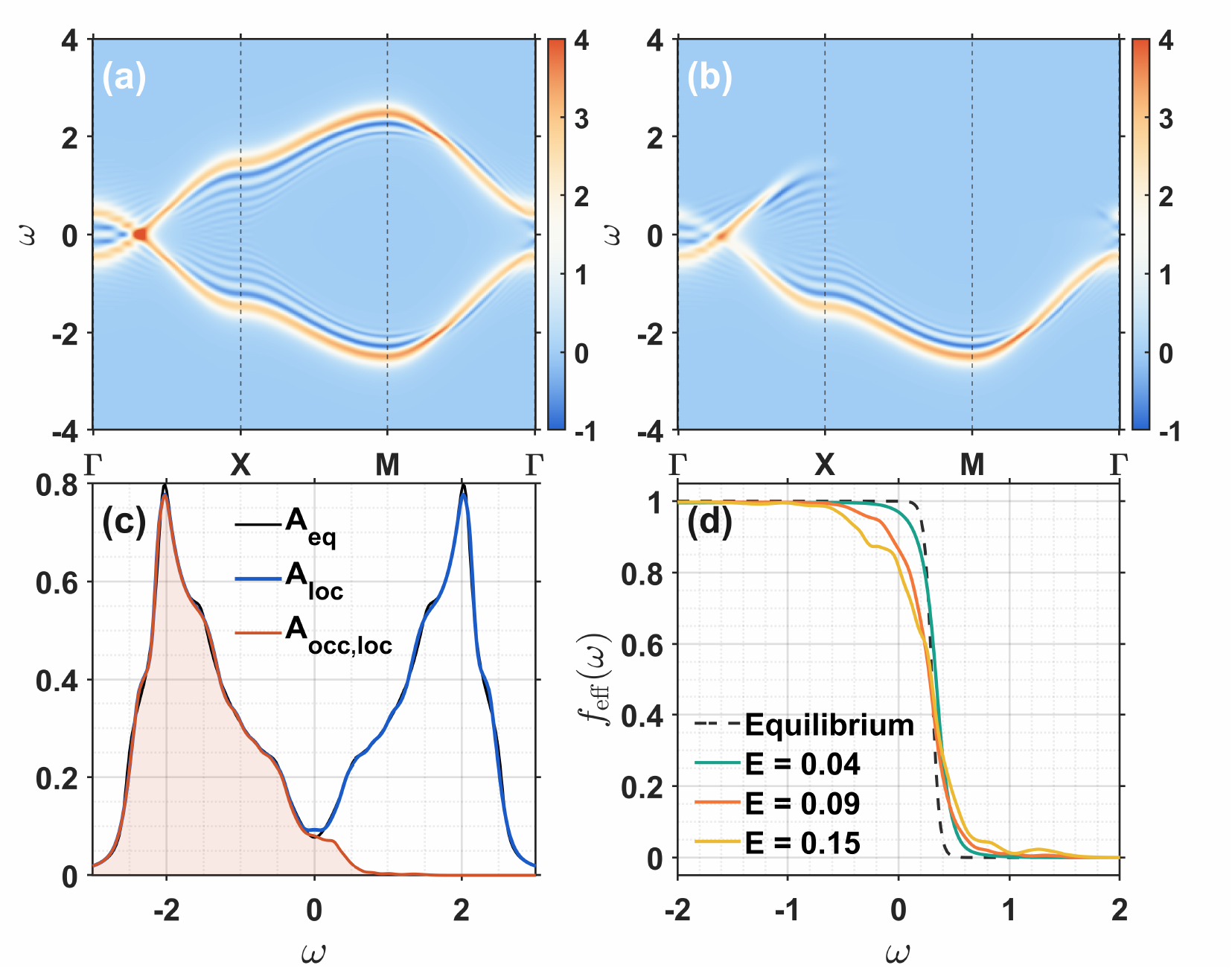}
\caption{
(a) Spectral function $\mathrm{Tr}\,A(\omega,\kappa_{\mathrm{path}})$ and (b) occupied spectral function $\mathrm{Tr}\,A_{\mathrm{occ}}(\omega,\kappa_{\mathrm{path}})$ along the high-symmetry path $\Gamma$--$X$--$M$--$\Gamma$, together with (c) the full local spectrum $A_{\mathrm{loc}}(\omega)$ and occupied local spectrum $A_{\mathrm{occ,loc}}(\omega)$, for the representative case $E=0.09$. 
(d) Effective local distribution function $f_{\mathrm{eff}}(\omega)$ in equilibrium and under driving fields $E=0.04$, $0.09$, and $0.15$.
}
\label{fig_3}
\end{figure}
So far, we have mainly used the constant-$\Gamma$ reservoir model as the simplest
example of a featureless reservoir-induced relaxation channel.
Material-specific dissipation processes generally introduce
additional frequency, momentum, and orbital dependence into
the self-energies. Such structures can be incorporated within the
NEGF framework, as illustrated by the following frequency-dependent bath,
electron-phonon, and interaction calculations, although the numerical
solution becomes substantially more demanding. The corresponding
formulations are described in the Supplemental Material~\cite{SM}.
Equilibrium spectroscopy and transport measurements can constrain the
retarded self-energy and characteristic relaxation scales, but they do
not uniquely determine the nonequilibrium dissipation mechanisms under strong driving.

The more general implementation also gives access to frequency-resolved quantities.
Figures~\ref{fig_3}(a) and \ref{fig_3}(b) show the momentum-resolved spectral function and occupied spectral weight along a high-symmetry path for a Lorentzian fermionic bath with peak amplitude $\Gamma=0.1$ and half-width $W=4$, at $T=1/30$, $\mu=0.3$, and $E=0.09$.
The stripe-like structures along the $\Gamma$--$X$ path are a spectral signature of 
interband resonances. 
We note that the imaginary part of the momentum-resolved
lesser Green's function in the Wigner representation should be interpreted as a quasi-distribution rather than a positive-definite probability density.
It can therefore display negative values, while its frequency-integrated momentum distribution, shown in Fig.~\ref{fig_2}, and its momentum-integrated occupied spectral weight, shown in Fig.~\ref{fig_3}(c), remain positive.

We further extract an effective distribution function $f_{\mathrm{eff}}(\omega)=\mathrm{Tr}\,A_\text{loc,occ}(\omega)/\mathrm{Tr}\,A_\text{loc}(\omega)$ from the local occupied spectral weight and the local spectral function in Fig.~\ref{fig_3}(c).
The result is shown in Fig.~\ref{fig_3}(d).
At weak field, for example $E=0.04$, the distribution already deviates from the equilibrium Fermi function but remains smooth.
At stronger fields, pronounced wiggles and a clear left-right asymmetry appear.
The detailed form of these structures depends on the band dispersion, but their emergence signals the breakdown of the perturbative regime and the formation of a strongly driven nonequilibrium steady state.

\begin{figure}[tbp]
    \centering
    \includegraphics[width=\columnwidth]{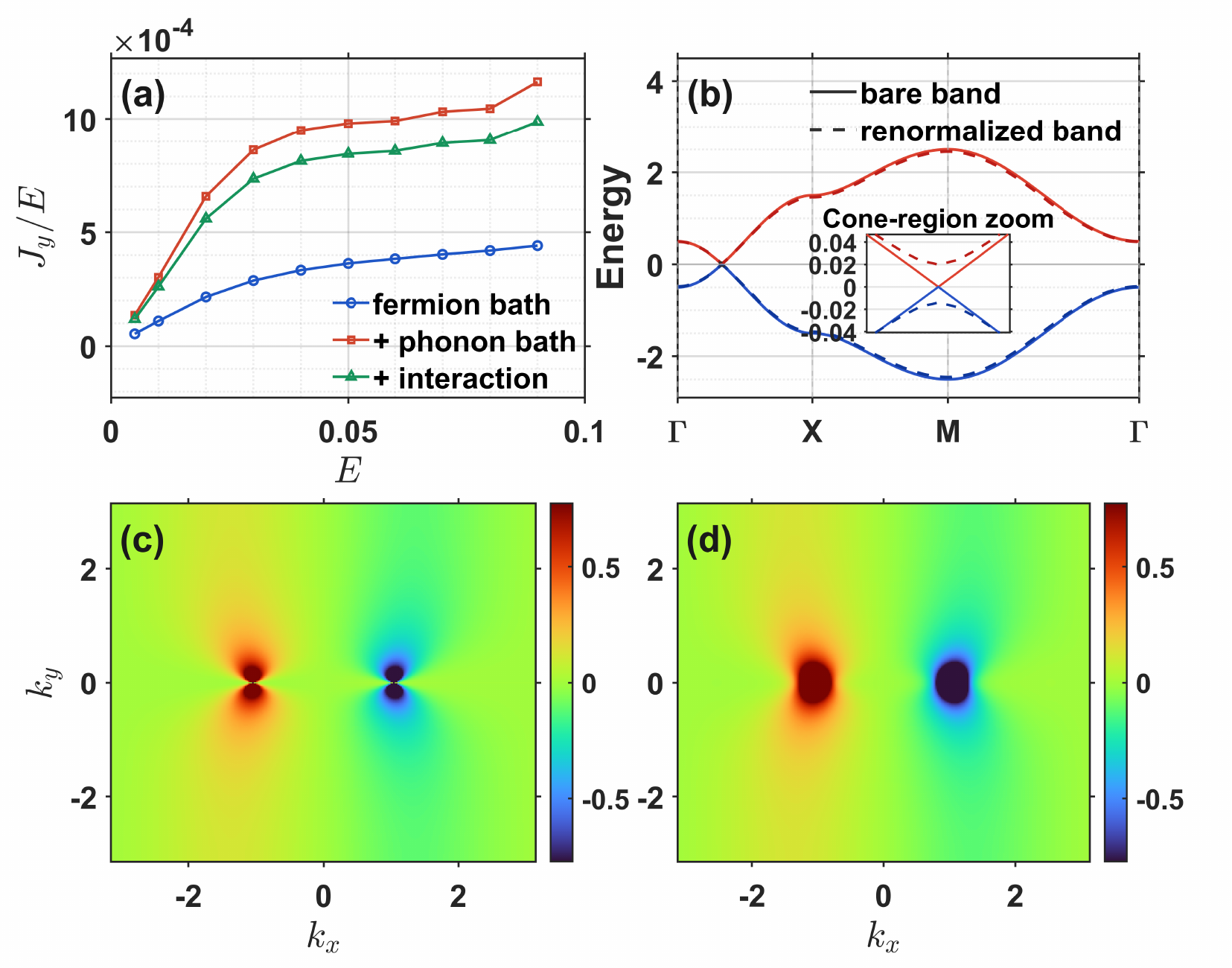}
\caption{
(a) Field-normalized transverse current $J_y/E$ as a function of the driving field $E$ for three cases: coupling only to a fermionic bath, coupling to both fermionic and phononic baths, and the interacting case. 
(b) Bare and phonon-renormalized band dispersions along the high-symmetry path $\Gamma$--$X$--$M$--$\Gamma$, with an inset showing a magnified view of the 
nodal region to highlight the gap opened by the real part of the self-energy $\Sigma_{\text{ph}}$. 
(c) Bare and (d) phonon-renormalized Berry curvature of the upper band across the Brillouin zone.
}
\label{fig_4}
\end{figure}

Finally, we examine how electron-phonon coupling and electronic
interactions modify the Hall response. Starting from the Lorentzian
fermionic-bath setup, we include either coupling to a phonon mode with
$\omega_0=1$, treated in the Migdal approximation, or a local Hubbard
interaction with $U=0.8$, treated within dynamical mean-field theory
using boldified second-order perturbation theory~\cite{Aoki2014}. As
shown in Fig.~\ref{fig_4}(a), the overall field dependence remains
similar in all three cases, demonstrating the robustness of the
nonperturbative mechanisms identified above. Both self-energies,
however, substantially enhance the magnitude of $J_y/E$.

This enhancement is governed mainly by the real part of the
self-energy, which renormalizes the band structure and Berry curvature,
rather than by the additional damping associated with its imaginary
part. The quasiparticle dispersion extracted from the equilibrium
electron-phonon self-energy, shown in Fig.~\ref{fig_4}(b), exhibits a
slight band flattening and a small gap near the nodal point. Despite
being modest on the scale of the bandwidth, these changes substantially
enhance the Berry-curvature hot spots, as shown in
Figs.~\ref{fig_4}(c) and \ref{fig_4}(d). Consistently, an effective
noninteracting model with
$D_y^{\rm eff}=D_y-\operatorname{Re}\Sigma_z^R(0)$ reproduces the Hall
current of the full electron-phonon calculation within a few percent
for $E\lesssim0.1$~\cite{SM}.

In summary, we studied the Berry-curvature-induced nonlinear Hall transport
in the nonperturbative regime and in the presence of electron correlations by employing
a nonequilibrium steady-state Green's function approach 
for dc-field-driven lattice systems.
The method treats the electric field nonperturbatively and gives access to the momentum- and frequency-resolved nonequilibrium electronic structure.
For a two-band model, we showed that the Berry-curvature-dipole picture is controlled only at sufficiently large doping and weak fields.
It breaks down when the driven Fermi-contour edge reaches the nodal point Berry-curvature hot spots, or when interband charge transfer becomes important in the semimetallic regime.
The same analysis applied to a $C_4\mathcal{T}$-symmetric model with a vanishing Berry-curvature dipole reproduces the expected quadrupolar weak-field scaling and its breakdown near the Berry-curvature hot spots~\cite{SM}, supporting the broader applicability of our framework and conclusions.
We also clarified that the constant-relaxation-time approximation and the constant-$\Gamma$ reservoir model can predict different crossover fields.
In the nonperturbative regimes, Berry-curvature-weighted occupation formulas no longer reliably reproduce the full Hall current, which is instead obtained directly within the NEGF framework.
The formalism also makes it possible to incorporate dissipation, interactions, and electron-phonon coupling through self-energies, going beyond the constant relaxation-time approximation.
Our DMFT calculations showed that interactions and electron-phonon coupling can substantially change the Hall response, not only through additional damping but also through band renormalizations and gap openings, and the resulting modifications of Berry-curvature hot spots.
These results demonstrate a flexible numerical framework for nonperturbative nonequilibrium nonlinear Hall phenomena and offer guidance for strong-field transport and spectroscopic experiments.

{\it Acknowledgments}
The calculations were run on the Beo06 cluster at the University of Fribourg. This work was supported by the Swiss National Science Foundation via NCCR Marvel and Grant No.~2000-1-240023, by the German Research Foundation through research unit QUAST (FOR 5249), and the Austrian Science Fund via Grant 10.55776/I5868 (QUAST). 
Work at Rice University has been supported by the National Science Foundation under Grant No. DMR-2220603 (L.C., S.S. and Q.S), and by the Robert A. Welch Foundation Grant No. C-1411 (Q.S.) and the Vannevar Bush Faculty Fellowship ONR-VB N00014-23-1-2870 (Q.S.).
M. E. acknowledges discussions with Karsten Held. 
L.C., S. P., Q.S. and P. W. acknowledge the hospitality of the Aspen Center for Physics, which is supported by the National Science Foundation under Grant No. PHY-2210452 and a grant from the Simons Foundation (1161654, Troyer).
\bibliography{mybibtex}

\clearpage
\onecolumngrid
\begin{center}
{\large\bfseries Supplemental Material for\\
Nonperturbative Nonlinear Hall Effect in Nonequilibrium Steady States}
\end{center}
\twocolumngrid
\setcounter{secnumdepth}{1}
\setcounter{section}{0}
\setcounter{equation}{0}
\setcounter{figure}{0}
\setcounter{table}{0}
\renewcommand{\theequation}{S\arabic{equation}}
\renewcommand{\thefigure}{S\arabic{figure}}
\renewcommand{\thetable}{S\arabic{table}}
\renewcommand{\theHequation}{supp.\arabic{equation}}
\renewcommand{\theHfigure}{supp.\arabic{figure}}
\renewcommand{\theHtable}{supp.\arabic{table}}
\renewcommand{\theHsection}{supp.\arabic{section}}

\section{Steady-state formalism based on nonequilibrium Green's functions}

\subsubsection{Moyal product}

The nonlinear Hall effect under a constant electric field, and in the presence of dissipation, is a nonequilibrium phenomenon, which can be studied using nonequilibrium Green's functions. In related previous works, the Green's functions of nonequilibrium steady states were obtained by solving the Schwinger-Dyson equations self-consistently. Specifically, 
the retarded and lesser Green's functions are obtained from~\cite{Li2021,Yuta2018}
\begin{equation}
D \ast G^{R} = 1,
\qquad
D \equiv \left(G_0^{R}\right)^{-1} - \Sigma^{R},
\label{eq:SD_retarded}
\end{equation}
and
\begin{equation}
G^{<}=G^{R}\ast \Sigma^{<}\ast G^{A},
\label{eq:SD_lesser}
\end{equation}
where \(\ast\) denotes the ordinary convolution. 
For a 
homogeneous 
steady state, the spatial dependence can be diagonalized in momentum space, so that the remaining convolution is only in time. 
Unless otherwise stated, the Hamiltonian, Green's functions, and
self-energies appearing below are matrices in orbital space, and each product includes a matrix multiplication. Scalar terms are understood to multiply the identity matrix.

To reformulate the steady-state equations in a form suitable for systems with a constant electric field, it is useful to adopt the Wigner representation.  
As a simple example, consider the convolution of two two-time functions,
\begin{equation}
C(t_1,t_2)
=
\int dt_3\,
A(t_1,t_3)\,B(t_3,t_2).
\label{eq:convolution_t}
\end{equation}
We introduce the center-of-mass time and relative time,
\begin{equation}
T=\frac{t_1+t_2}{2},
\qquad
t=t_1-t_2,
\end{equation}
and define the Wigner transform of a two-time function by
\begin{equation}
A(T,\omega)
=
\int dt\,
e^{i\omega t}
A\!\left(T+\frac{t}{2},\,T-\frac{t}{2}\right).
\label{eq:wigner_def}
\end{equation}
In this representation, the convolution in Eq.~\eqref{eq:convolution_t} is mapped to the Moyal, or star, product, denoted as $\star$~\cite{groenewold1946principles,moyal1949quantum},
\begin{equation}
\begin{aligned}
C(T,\omega) &= (A \star B)(T,\omega) \\
&= A(T,\omega) \exp \left[ \frac{i}{2} \left( \overleftarrow{\partial_\omega}\overrightarrow{\partial_T} - \overleftarrow{\partial_T}\overrightarrow{\partial_\omega} \right) \right] B(T,\omega).
\end{aligned}
\label{eq:moyal_product}
\end{equation}
Left (right) arrows above the derivative operators indicate that they act on the left (right). Thus, even in the absence of external fields, the Wigner representation converts the original time convolution into a differential product structure in the \((T,\omega)\) representation.

\subsubsection{Gauge-covariant formulation}

This formulation becomes especially useful when extending the theory to a system driven by a constant electric field \cite{Aron2012}. 
In that case, it is no longer sufficient to work only with the time variables, because the electric field couples temporal and spatial degrees of freedom through the electromagnetic four-potential. 
It is therefore natural to generalize the Wigner representation from the two-time form discussed above to the full four-dimensional phase-space representation.

For functions depending on the center-of-mass coordinate \(X^\mu\) and conjugate four-momentum \(k^\mu\), the ordinary Moyal product takes the form
\begin{equation}
A(X,k)
\exp\!\left[
\frac{i}{2}
\left(
\overleftarrow{\partial_{k^{\mu}}}\overrightarrow{\partial_{X_{\mu}}}
-
\overleftarrow{\partial_{X^{\mu}}}\overrightarrow{\partial_{k_{\mu}}}
\right)
\right]
B(X,k).
\label{eq:moyal_4d}
\end{equation}
In the presence of an electromagnetic field, however, the Wigner representation must be reformulated in a gauge-covariant manner.
This construction is standard in the derivation of quantum kinetic theory, where the usual Wigner transform is replaced by its gauge-invariant counterpart, equivalently expressed in terms of the kinetic momentum.
To this end, one introduces the gauge-covariant four-momentum
\begin{equation}
\kappa^\mu = k^\mu - q A^\mu(X),
\label{eq:gauge_covariant_momentum}
\end{equation}
with $A^\mu(X)$ the electromagnetic four-potential,
and the Wigner Green's function
$ \widetilde G(X,\boldsymbol{\kappa})=G(X,\boldsymbol{\kappa}+q\boldsymbol{A})$.
The Schwinger-Dyson equations then take the form of gauge-covariant Moyal star products in the $(X,\kappa)$ variables~\cite{Aron2012,onoda2006theory,JKFreericks_2006,Bertoncini1991}.

\begin{align}
(A \star B)(X,\omega,\boldsymbol{\kappa})
&=
A(X,\omega,\boldsymbol{\kappa})
\exp\Bigg[
\frac{i}{2}
\left(
\overleftarrow{\partial_{\kappa^{\mu}}}\overrightarrow{\partial_{X_{\mu}}}
-
\overleftarrow{\partial_{X^{\mu}}}\overrightarrow{\partial_{\kappa_{\mu}}}
\right)
\nonumber\\
&
+\frac{i}{2} q \mathbf{E}\!\cdot\!
\left(
\overleftarrow{\partial_\omega}\overrightarrow{\nabla_{\boldsymbol{\kappa}}}
-
\overleftarrow{\nabla_{\boldsymbol{\kappa}}}\overrightarrow{\partial_\omega}
\right)
\Bigg]
B(X,\omega,\boldsymbol{\kappa}),
\label{eq:star_product_EM}
\end{align}
where we have used \(\kappa^0=\omega\) and \(\boldsymbol{\kappa}=(\kappa^1,\kappa^2,\kappa^3)\). 
The second term in the exponent is the field-induced contribution, which mixes frequency and momentum derivatives and reflects the gauge-covariant structure of the steady-state problem under a dc field.

For a spatially uniform nonequilibrium steady state driven by a constant electric field, all quantities are independent of the center-of-mass time and space \(X\), and the first term in the exponent of Eq.~\eqref{eq:star_product_EM} drops out. 
The star product then reduces to
\begin{align}
&(A \star B)(\omega,\boldsymbol{\kappa})
=\nonumber\\
& \quad A(\omega,\boldsymbol{\kappa})
\exp\!\left[
\frac{i}{2} q \mathbf{E}\cdot
\left(
\overleftarrow{\partial_\omega}\overrightarrow{\nabla_{\boldsymbol{\kappa}}}
-
\overleftarrow{\nabla_{\boldsymbol{\kappa}}}\overrightarrow{\partial_\omega}
\right)
\right]
B(\omega,\boldsymbol{\kappa}),
\label{eq:star_product_dc}
\end{align}
and Schwinger-Dyson equations in a constant electric field are written in terms of this gauge-covariant star product rather than the ordinary convolution.

The gauge-covariant star product in Eq.~\eqref{eq:star_product_dc} can be implemented either in the frequency domain or in the time domain. 
For sufficiently strong electric fields, the frequency-domain formulation is advantageous, and has been used in earlier works~\cite{Aron2012,Aron2012b,Aron2013}. 
In contrast, for weaker electric fields the time-domain formulation is generally more convenient from the numerical point of view. In the present study of the nonlinear Hall effect, we have examined both formulations. 
For the typical field strengths considered here, the time-domain approach exhibits faster convergence. 
We therefore focus in the following on the time-domain implementation of the star product and the associated Schwinger-Dyson equations.

For the time-domain solution of the Dyson equations with the gauge-covariant star product, we employ two different implementations. 
In one case, the self-energy is fully frequency dependent, whereas the other employs the constant-$\Gamma$ reservoir model. 
The latter case leads to substantial simplifications and can be solved much more efficiently numerically. 
In practice, a typical calculation within this simplified scheme requires only a few seconds on a 128-core computing node, whereas the frequency-dependent case may require several hours. 
We first describe the constant-$\Gamma$ reservoir model. 

\subsubsection{Constant-$\Gamma$ reservoir model}

We start from Eq.~\eqref{eq:SD_retarded}, with the ordinary convolution replaced by the gauge-covariant star product. 
For this model, the orbital-independent retarded self-energy is
$\Sigma^R=-i\Gamma$, we obtain
\begin{equation}
\Big[
\omega+i\Gamma-H(\boldsymbol{\kappa})
\Big]\star
\widetilde G^R(\omega,\boldsymbol{\kappa})
=1.
\label{eq:Dyson_star_freq}
\end{equation}

Applying Eq.~\eqref{eq:star_product_dc} to Eq.~\eqref{eq:Dyson_star_freq}, one finds
\begin{equation}
\left[
\omega+i\Gamma
+\frac{i}{2}\,\mathbf{E}\!\cdot\!\nabla_{\boldsymbol{\kappa}}
-H\left(\boldsymbol{\kappa}-\frac{i}{2}\mathbf{E}\,\partial_\omega\right)
\right]
\widetilde G^{R}(\omega,\boldsymbol{\kappa})
=1,
\label{eq:Dyson_bopp}
\end{equation}
where we have absorbed the charge \(q\) into the definition of \(\mathbf{E}\) for notational simplicity.

We next perform a Fourier transform with respect to \(\omega\), using
\begin{equation}
\omega\,\widetilde G(\omega,\boldsymbol{\kappa})
\;\longleftrightarrow\;
i\partial_\tau \widetilde G(\tau,\boldsymbol{\kappa}),
\quad
\partial_\omega \widetilde G(\omega,\boldsymbol{\kappa})
\;\longleftrightarrow\;
i\tau\,\widetilde G(\tau,\boldsymbol{\kappa}).
\label{eq:Fourier_rules}
\end{equation}
Equation~\eqref{eq:Dyson_bopp} is then transformed into
\begin{equation}
\left[
i\partial_\tau
+i\Gamma
+\frac{i}{2}\,\mathbf{E}\!\cdot\!\nabla_{\boldsymbol{\kappa}}
-H\left(\boldsymbol{\kappa}+\frac{1}{2}\mathbf{E}\tau\right)
\right]
\widetilde G^{R}(\tau,\boldsymbol{\kappa})
=
\delta(\tau).
\label{eq:Dyson_tau_kappa_with_advection}
\end{equation}

To eliminate the drift term, we introduce the comoving momentum variable
\begin{equation}
\boldsymbol{\kappa}' \equiv \boldsymbol{\kappa} - \frac{1}{2}\mathbf{E}\tau.
\label{eq:comoving_kappa}
\end{equation}
By the chain rule,
\begin{equation}
\partial_\tau\big|_{\boldsymbol{\kappa}}
=
\partial_\tau\big|_{\boldsymbol{\kappa}'}
-\frac{1}{2}\mathbf{E}\!\cdot\!\nabla_{\boldsymbol{\kappa}'},
\qquad
\nabla_{\boldsymbol{\kappa}}\big|_{\tau}
=
\nabla_{\boldsymbol{\kappa}'},
\label{eq:chain_rule}
\end{equation}
so that
\begin{equation}
i\partial_\tau\big|_{\boldsymbol{\kappa}}
+\frac{i}{2}\mathbf{E}\!\cdot\!\nabla_{\boldsymbol{\kappa}}
=
i\partial_\tau\big|_{\boldsymbol{\kappa}'}.
\label{eq:advection_cancel}
\end{equation}
Defining \( {G'}^{R}(\tau,\boldsymbol{\kappa}') \equiv\widetilde G^{R}(\tau,\boldsymbol{\kappa})\), Eq.~\eqref{eq:Dyson_tau_kappa_with_advection} reduces to
\begin{equation}
\left[
i\partial_\tau
+i\Gamma
-H\left(\boldsymbol{\kappa}'+\mathbf{E}\tau\right)
\right]
{G'}^{R}(\tau,\boldsymbol{\kappa}')
=
\delta(\tau).
\label{eq:Dyson_tau_kappa_final}
\end{equation}

Since \( G'^{R}\) is a retarded Green's function, it satisfies
\begin{equation}
 G'^{R}(\tau,\boldsymbol{\kappa}')=0,
\qquad \tau<0,
\end{equation}
and the discontinuity at \(\tau=0\) implies
\begin{equation}
 G'^{R}(0^+,\boldsymbol{\kappa}')=-i.
\end{equation}

For $\tau>0$, Eq.~\eqref{eq:Dyson_tau_kappa_final} becomes
the homogeneous matrix differential equation
\begin{equation}
\left[
i\partial_\tau+i\Gamma
-
H\left(\boldsymbol{\kappa}'+\mathbf E\tau\right)
\right]
{G'}^R(\tau,\boldsymbol{\kappa}')
=0.
\end{equation}
Because the Hamiltonians at different points along the field-driven
trajectory do not generally commute, its solution is expressed as a
time-ordered exponential,
\begin{equation}
{G'}^R(\tau,\boldsymbol{\kappa}')
=
-i\theta(\tau)e^{-\Gamma\tau}
\mathcal T
\exp\left[
-i\int_0^\tau d\tau'\,
H\left(\boldsymbol{\kappa}'+\mathbf E\tau'\right)
\right].
\label{eq:G0R_tau_comoving}
\end{equation}
Here, $\mathcal T$ denotes chronological time ordering.

Transforming back to the original momentum variable gives
\begin{equation}
\widetilde G^R(\tau,\boldsymbol{\kappa})
=
-i\theta(\tau)e^{-\Gamma\tau}
\mathcal T
\exp\left[
-i\int_{-\tau/2}^{\tau/2}d\tau'\,
H\left(\boldsymbol{\kappa}+\mathbf E\tau'\right)
\right].
\label{eq:G0R_tau_correct}
\end{equation}
Numerically, the time-ordered exponential is evaluated by dividing
the propagation interval into short time steps and multiplying the
corresponding short-time matrix propagators in chronological order.
For the centered-time form in Eq.~\eqref{eq:G0R_tau_correct},
the propagation starts at $\tau=0$, with each time step adding a
short-time propagator from the left and right for the positive- and
negative-time branches, respectively.
This procedure retains the noncommutativity of the Hamiltonians at
different times.

We next consider the calculation of the lesser Green's function in Eq.~\eqref{eq:SD_lesser}, with the ordinary convolution replaced by the star product. 
Using the Fourier transform together with Eq.~\eqref{eq:Fourier_rules}, the dc star product in Eq.~\eqref{eq:star_product_dc} can be rewritten in the time domain as
\begin{equation}
\begin{aligned}
(A\star B)(\tau,\boldsymbol{\kappa})
=
\int d\tau_1\,
&A\!\left(
\tau_1,
\boldsymbol{\kappa}
+\frac{\mathbf{E}}{2}(\tau-\tau_1)
\right)
\\
\times\,
&B\!\left(
\tau-\tau_1,
\boldsymbol{\kappa}
-\frac{\mathbf{E}}{2}\tau_1
\right).
\end{aligned}
\label{eq:star_time_convolution}
\end{equation}
Since the bath self-energy 
$\Sigma$ 
is independent of \(\boldsymbol{\kappa}\), the star product of the first two factors in Eq.~\eqref{eq:SD_lesser} becomes
\begin{equation}
\begin{aligned}
(\widetilde G^R\star \Sigma^{<})(\tau,\boldsymbol{\kappa})
=
\int d\tau_1 \,
&\widetilde G^R\!\left(
\tau_1,
\boldsymbol{\kappa}
+\frac{\mathbf{E}}{2}(\tau-\tau_1)
\right)
\\
\times\,
&\Sigma^{<}(\tau-\tau_1).
\end{aligned}
\label{eq:sigma_right_star}
\end{equation}
Applying the star product once more then yields 
\begin{align}
&(\widetilde G^R\star \Sigma^{<}\star \widetilde G^A)(\tau,\boldsymbol{\kappa})
=
\int d\tau_1 d\tau_2 \,
\widetilde G^R\!\left(
\tau_1,
\boldsymbol{\kappa}
+\frac{\mathbf{E}}{2}(\tau-\tau_1)
\right)
\nonumber\\
&\hspace{1cm}\times\,\Sigma^{<}(\tau_2-\tau_1)\,
\widetilde G^A\!\left(
\tau-\tau_2,
\boldsymbol{\kappa}
-\frac{\mathbf{E}}{2}\tau_2
\right).\label{eq:sigma_left_star}
\end{align}

Since the current requires only the equal-time lesser Green's
function, it is not necessary to evaluate the complete
frequency-resolved lesser Green's function. We therefore compute the
momentum-resolved density matrix
\begin{equation}
\widetilde n(\boldsymbol{\kappa})
=
-i\widetilde G^<(0,\boldsymbol{\kappa}).
\label{eq:occupation_def}
\end{equation}
Substituting Eq.~\eqref{eq:sigma_left_star} with \(\tau=0\), we obtain
\begin{equation}
\begin{aligned}
\widetilde n(\boldsymbol{\kappa})
=
-i\int d\tau_1 d\tau_2 \,
&\widetilde G^R\!\left(
\tau_1,
\boldsymbol{\kappa}-\frac{\mathbf{E}}{2}\tau_1
\right)
\,\Sigma^{<}(\tau_2-\tau_1)
\\
\times\,
&\widetilde G^A\!\left(
-\tau_2,
\boldsymbol{\kappa}-\frac{\mathbf{E}}{2}\tau_2
\right).
\end{aligned}
\label{eq:occupation_intermediate}
\end{equation}
Using the steady-state relation
\begin{equation}
\widetilde G^A(-\tau,\boldsymbol{\kappa})
=
\left[
\widetilde G^R(\tau,\boldsymbol{\kappa})
\right]^\dagger.
\label{eq:advanced_retarded_relation}
\end{equation}
this expression can be rewritten as
\begin{align}
&\widetilde n(\boldsymbol{\kappa})
=
-i\int d\tau_1 d\tau_2 \,
\widetilde G^R\!\left(
\tau_1,
\boldsymbol{\kappa}-\frac{\mathbf{E}}{2}\tau_1
\right)
\,\Sigma^{<}(\tau_2-\tau_1)
\nonumber\\
&\hspace{2cm}\times\,
\left[\widetilde G^{R}\!\left(
\tau_2,
\boldsymbol{\kappa}-\frac{\mathbf{E}}{2}\tau_2
\right)\right]^\dagger.\label{eq:occupation_with_GR}
\end{align}

For the constant bath considered here, the lesser self-energy takes the form
\begin{equation}
\Sigma^{<}(\tau_2-\tau_1)
=
2i\Gamma
\int\frac{d\omega}{2\pi}\,
e^{-i\omega(\tau_2-\tau_1)} f(\omega),
\label{eq:Sigma_less_constbath}
\end{equation}
where \(f(\omega)\) is the bath distribution function. 
Substituting Eq.~\eqref{eq:Sigma_less_constbath} into
Eq.~\eqref{eq:occupation_with_GR}, we obtain
\begin{align}
\widetilde n(\boldsymbol{\kappa})
=
\Gamma\int\frac{d\omega}{\pi}f(\omega)
&\left[
\int_0^\infty d\tau_1\,
\widetilde G^R\left(
\tau_1,
\boldsymbol{\kappa}-\frac{\mathbf E}{2}\tau_1
\right)e^{i\omega\tau_1}
\right]
\nonumber\\
{}\times&
\left[
\int_0^\infty d\tau_2\,
\widetilde G^R\left(
\tau_2,
\boldsymbol{\kappa}-\frac{\mathbf E}{2}\tau_2
\right)e^{i\omega\tau_2}
\right]^\dagger .
\label{eq:occupation_factorized}
\end{align}
This motivates the definition of an effective spectral amplitude,
\begin{equation}
\mathcal{G}(\omega,\boldsymbol{\kappa})
=
\int_0^{\infty} d\tau\,
\widetilde G^R\!\left(
\tau,
\boldsymbol{\kappa}-\frac{\mathbf{E}}{2}\tau
\right)
e^{i\omega\tau},
\label{eq:effective_spectral_amplitude}
\end{equation}
in terms of which the occupation can be written in the compact form
\begin{equation}
\widetilde n(\boldsymbol{\kappa})
=
\Gamma\int\frac{d\omega}{\pi}\,
f(\omega)\,
\mathcal G(\omega,\boldsymbol{\kappa})
\mathcal G^\dagger(\omega,\boldsymbol{\kappa}).
\label{eq:occupation_final}
\end{equation}

\begin{figure}[!tbp]
    \centering
    \includegraphics[width=\columnwidth]{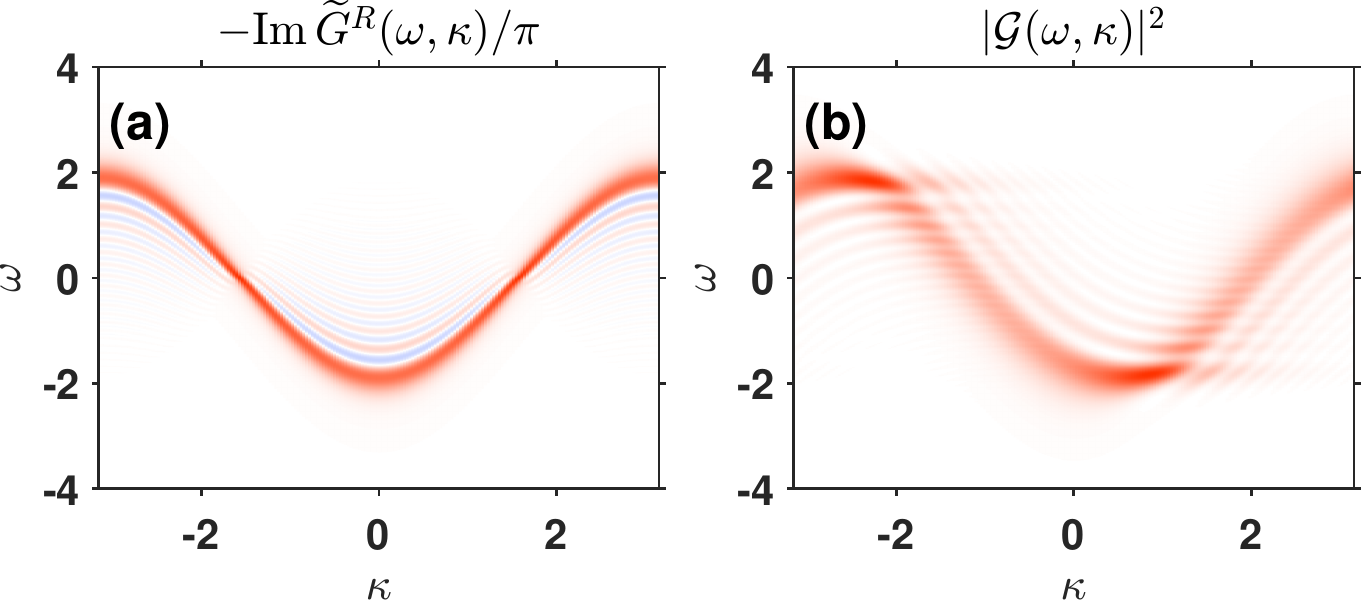}
\caption{
Momentum- and frequency-resolved spectra for a single-band
one-dimensional tight-binding model with $E=\Gamma=0.1$.
(a) Wigner spectral function $-\mathrm{Im}\,\widetilde G^R(\omega,\kappa)/\pi$.
(b) Field-dressed weight $|\mathcal{G}(\omega,\kappa)|^2$.
}
\label{fig:SM_effective_spectral}
\end{figure}

To clarify the difference between Eq.~\eqref{eq:effective_spectral_amplitude} and the Wigner Green's function in Eq.~\eqref{eq:G0R_tau_correct}, we compare them for a one-dimensional single-band tight-binding model with $\Gamma=E=0.1$ in Fig.~\ref{fig:SM_effective_spectral}. In this case, $\mathcal G\mathcal G^\dagger$ reduces to $|\mathcal G|^2$.
Both the Wigner spectral function $-\mathrm{Im}\,\widetilde G^R(\omega,\kappa)/\pi$ and the effective spectral weight $|\mathcal{G}(\omega,\kappa)|^2$ develop field-induced oscillatory structures.
The Wigner spectral function remains symmetric with respect to $\kappa=0$ in this model, but can take locally negative values because it is a quasi-distribution in Wigner space.
In contrast, $|\mathcal{G}(\omega,\kappa)|^2$ is positive by construction and becomes asymmetric with respect to $\kappa=0$, reflecting how the field-driven momentum-space structure affects the occupation.
It is therefore more directly connected to the nonequilibrium momentum distribution in Eq.~\eqref{eq:occupation_final}.

Once the density matrix has been obtained, the current is evaluated as
\begin{equation}
\mathbf j
=
q\int\frac{d^d\kappa}{(2\pi)^d}
\operatorname{Tr}\left[
\mathbf v(\boldsymbol{\kappa})
\widetilde n(\boldsymbol{\kappa})
\right],
\label{eq:current_from_occupation}
\end{equation}
with the velocity operator
\begin{equation}
\mathbf v(\boldsymbol{\kappa})
=
\frac{1}{\hbar}
\nabla_{\boldsymbol{\kappa}}
H(\boldsymbol{\kappa}).
\label{eq:velocity_def}
\end{equation}

\subsubsection{Frequency-dependent self-energy}

We next turn to the more general case of a frequency-dependent self-energy. 
This formulation is required when considering a frequency-dependent bath, phonon coupling, or interaction-induced self-energy corrections. 
In this case, the simplifications used in the constant-damping approximation are no longer applicable, and the calculation must be carried out in terms of explicit multiple time integrals. 
In particular, Eq.~\eqref{eq:SD_retarded} for the retarded component becomes an integro-differential equation in the time domain. 
We therefore rewrite it in the iterative form
\begin{equation}
\widetilde G^R(\tau,\boldsymbol{\kappa})
=
\widetilde G_0^R(\tau,\boldsymbol{\kappa})
+
\bigl(
\widetilde G_0^R \star \Delta\Sigma^R \star \widetilde G^R
\bigr)(\tau,\boldsymbol{\kappa}),
\label{eq:Dyson_star_time}
\end{equation}
where \(\widetilde G_0^R\) denotes the field-dressed propagator in the absence of the frequency-dependent correction \(\Delta\Sigma^R\). 
It is obtained from Eq.~\eqref{eq:G0R_tau_correct} by setting \(\Gamma=0\). 
Here \(\Delta\Sigma^R\) represents the self-energy correction arising from a frequency-dependent bath, phonons, or interactions.

To evaluate the triple product in Eq.~\eqref{eq:Dyson_star_time}, we proceed in two steps and define
\begin{equation}
M(\tau,\boldsymbol{\kappa})
\equiv
(\Delta\Sigma^R \star \widetilde G^R)(\tau,\boldsymbol{\kappa}),
\label{eq:M_def}
\end{equation}
and
\begin{equation}
\mathrm{corr}(\tau,\boldsymbol{\kappa})
\equiv
(\widetilde G_0^R \star M)(\tau,\boldsymbol{\kappa}),
\label{eq:corr_def}
\end{equation}
so that the Dyson equation becomes
\begin{equation}
\widetilde G^R(\tau,\boldsymbol{\kappa})
=
\widetilde G_0^R(\tau,\boldsymbol{\kappa})
+
\mathrm{corr}(\tau,\boldsymbol{\kappa}).
\label{eq:Dyson_two_step}
\end{equation}

Since \(\Delta\Sigma^R\) is assumed to be independent of \(\boldsymbol{\kappa}\), the first star product can be written, using Eq.~\eqref{eq:star_time_convolution}, as
\begin{equation}
M(\tau,\boldsymbol{\kappa})
=
\int_0^\tau d\tau_1\;
\Delta\Sigma^R(\tau_1)\,
\widetilde G^R\!\left(
\tau-\tau_1,\,
\boldsymbol{\kappa}
-\frac{\mathbf{E}}{2}\tau_1
\right).
\label{eq:M_def_time}
\end{equation}
The field-induced shift is only along the electric-field direction. 
We therefore perform the Fourier expansion along this direction. 
Introducing
\begin{equation}
\kappa_\parallel=\hat{\mathbf E}\cdot\boldsymbol{\kappa},
\qquad
\hat{\mathbf E}=\frac{\mathbf E}{|\mathbf E|},
\end{equation}
and denoting the remaining momentum components by \(\boldsymbol{\kappa}_\perp\), we expand
\begin{equation}
\widetilde G^R(\tau,\boldsymbol{\kappa})
=
\sum_n e^{in\kappa_\parallel}
\widetilde G_n^R(\tau,\boldsymbol{\kappa}_\perp),
\end{equation}
and similarly
\begin{equation}
M(\tau,\boldsymbol{\kappa})
=
\sum_n e^{in\kappa_\parallel}
M_n(\tau,\boldsymbol{\kappa}_\perp).
\label{eq:kappa_harmonics}
\end{equation}
Then
\begin{equation}
\begin{aligned}
&\widetilde G^R\!\left(
\tau-\tau_1,\,
\boldsymbol{\kappa}
-\frac{\mathbf{E}}{2}\tau_1
\right)
\\
&\quad =
\sum_n e^{in\kappa_\parallel}
e^{-in|\mathbf E|\tau_1/2}
\widetilde G_n^R(\tau-\tau_1,\boldsymbol{\kappa}_\perp),
\end{aligned}
\label{eq:G_shift_harmonics}
\end{equation}
so that each harmonic satisfies an independent causal convolution,
\begin{equation}
M_n(\tau,\boldsymbol{\kappa}_\perp)
=
\int_0^\tau d\tau_1\;
K_n(\tau_1)
\widetilde G_n^R(\tau-\tau_1,\boldsymbol{\kappa}_\perp),
\label{eq:Mn_conv}
\end{equation}
with
\begin{equation}
K_n(\tau_1)
\equiv
\Delta\Sigma^R(\tau_1)
e^{-in|\mathbf E|\tau_1/2}.
\label{eq:Kn_def}
\end{equation}
Thus, the first star product is reduced to an ordinary causal convolution in time for each harmonic channel along the field direction. 
On a uniform grid \(\tau_m=m\Delta\tau\), this convolution becomes
\begin{equation}
(M_n)_m
=
\Delta\tau
\sum_{m'=0}^{m}
(K_n)_{m'}\,
(\widetilde G_n)_{m-m'}.
\label{eq:Mn_discrete}
\end{equation}
Numerically, this convolution is evaluated using a zero-padded fast Fourier transform (FFT), and \(M(\tau,\boldsymbol{\kappa})\) is then reconstructed by an inverse FFT along the \(\kappa_\parallel\) direction.

The second star product is more involved, because both functions depend on \(\tau\) and \(\boldsymbol{\kappa}\). 
Consequently, a fixed harmonic channel of the final result receives contributions from different combinations of harmonic channels of \(\widetilde G_0^R\) and \(M\), and the product no longer reduces to independent one-dimensional convolutions which can be calculated using FFT. 
Explicitly, the correction term is
\begin{equation}
\begin{aligned}
\mathrm{corr}(\tau,\boldsymbol{\kappa})
=
\int_0^\tau d\tau_2\;
&\widetilde G_0^R\!\left(
\tau_2,\,
\boldsymbol{\kappa}
+\frac{\mathbf{E}}{2}(\tau-\tau_2)
\right)
\\
\times\,
&M\!\left(
\tau-\tau_2,\,
\boldsymbol{\kappa}
-\frac{\mathbf{E}}{2}\tau_2
\right).
\end{aligned}
\label{eq:corr_def_time}
\end{equation}
We evaluate this integral by direct summation on the discrete time grid. 
The momentum shifts are implemented spectrally rather than by interpolation, using the same idea as in Eq.~\eqref{eq:G_shift_harmonics}. 
For a periodic function along the field direction,
\begin{equation}
f(\kappa_\parallel)
=
\sum_n e^{in\kappa_\parallel} f_n,
\end{equation}
one has
\begin{equation}
f(\kappa_\parallel+s)
=
\sum_n e^{in\kappa_\parallel}
\bigl(e^{ins}f_n\bigr).
\label{eq:kappa_shift_phase}
\end{equation}
Thus each required shift along \(\mathbf E\) is implemented as
\begin{equation}
\{f(\kappa_{\parallel,j})\}_j
\xrightarrow{\ \mathrm{FFT}\ }
\{f_n\}_n
\longrightarrow
\{e^{ins}f_n\}_n
\xrightarrow{\ \mathrm{IFFT}\ }
\{f(\kappa_{\parallel,j}+s)\}_j .
\end{equation}
This procedure preserves periodicity along the field direction and avoids interpolation errors.

The Dyson equation is then solved iteratively, starting from the initial guess
\begin{equation}
\widetilde G^R(\tau,\boldsymbol{\kappa})
=
\widetilde G_0^R(\tau,\boldsymbol{\kappa}),
\end{equation}
and updating the solution through Eq.~\eqref{eq:Dyson_two_step}. 
In practice, linear mixing is applied after each iteration to improve numerical stability and convergence.

The lesser Green's function with a frequency-dependent self-energy can be treated in a closely analogous manner. 
For the lesser component, the star-product expression in Eq.~\eqref{eq:sigma_left_star} contains the same type of triple product as in the retarded Dyson equation. 
We therefore evaluate it in two steps, following the same strategy as for the retarded component. 
We first define
\begin{equation}
L(\tau,\boldsymbol{\kappa})
\equiv
(\Sigma^< \star \widetilde G^A)(\tau,\boldsymbol{\kappa}),
\label{eq:L_def}
\end{equation}
and then compute
\begin{equation}
\widetilde G^<(\tau,\boldsymbol{\kappa})
=
(\widetilde G^R \star L)(\tau,\boldsymbol{\kappa}).
\label{eq:Gless_two_step}
\end{equation}

The numerical strategy is the same as for the retarded component. 
Since \(\Sigma^<\) is assumed to be independent of \(\boldsymbol{\kappa}\), the first star product can again be reduced to independent convolutions in the \(\kappa_\parallel\)-harmonic representation and accelerated by FFT. 
The second star product is evaluated by direct summation, with the momentum shifts implemented spectrally as described above. 
The main difference from the retarded case is the time domain on which the functions are defined. 
Retarded and advanced quantities have causal support only on the positive and negative time branches, respectively, whereas \(\widetilde G^<(\tau,\boldsymbol{\kappa})\) and \(\Sigma^<(\tau)\) are defined on the full relative-time interval. 
Therefore, the numerical implementation of the lesser component must be carried out on the full time grid rather than only on the causal half axis.
\section{Weak-field comparison between the constant-relaxation-time approximation and the constant-$\Gamma$ reservoir model}
\label{sec:SM_tau_vs_Gamma}

In this section we clarify the difference between the commonly used constant-relaxation-time approximation and the constant-$\Gamma$ reservoir model used in our NEGF calculation.
Although the two descriptions are often identified in studies of linear transport, they lead to different momentum-space distortions of the distribution, which becomes important when estimating the onset of the nonperturbative regime.

For simplicity, we consider a one-dimensional single-band problem with the electric field applied along the momentum coordinate $\kappa$.
Starting from the effective spectral amplitude defined in Eq.~\eqref{eq:effective_spectral_amplitude} and the occupation formula in Eq.~\eqref{eq:occupation_final}, the weak-field expansion of the spectral amplitude gives
\begin{equation}
\mathcal{G}(\omega,\kappa)
=
\mathcal{G}_0(\omega,\kappa)
-
iE v_\kappa
\mathcal{G}_0^3(\omega,\kappa)
+
O(E^2),
\label{eq:SM_calG_weak_expand}
\end{equation}
where
\begin{equation}
\mathcal{G}_0(\omega,\kappa)
=
\frac{1}{\omega-\epsilon_\kappa+i\Gamma},
\qquad
v_\kappa=\partial_\kappa\epsilon_\kappa .
\end{equation}
Substituting Eq.~\eqref{eq:SM_calG_weak_expand} into Eq.~\eqref{eq:occupation_final}, the first-order correction to the occupation is
\begin{equation}
\delta \widetilde n_\Gamma(\kappa)
=
-4E\Gamma^2 v_\kappa
\int\frac{d\omega}{\pi}
f(\omega)
\frac{\omega-\epsilon_\kappa}
{\left[(\omega-\epsilon_\kappa)^2+\Gamma^2\right]^3}.
\label{eq:SM_delta_n_general}
\end{equation}

At zero temperature, \(f(\omega)=\theta(\mu-\omega)\).
Using integration by parts, Eq.~\eqref{eq:SM_delta_n_general} can be rewritten as
\begin{equation}
\delta \widetilde n_{\Gamma}(\kappa)
=
-E\Gamma^2 v_\kappa
\int\frac{d\omega}{\pi}
\frac{f'(\omega)}
{\left[(\omega-\epsilon_\kappa)^2+\Gamma^2\right]^2}.
\label{eq:SM_delta_n_fprime}
\end{equation}
Since \(f'(\omega)=-\delta(\omega-\mu)\), this gives
\begin{equation}
\delta \widetilde n_{\Gamma}(\kappa)
=
\frac{E\Gamma^2 v_\kappa}
{\pi\left[
(\mu-\epsilon_\kappa)^2+\Gamma^2
\right]^2}.
\label{eq:SM_delta_n_zeroT}
\end{equation}

The zero-field occupation obtained from Eq.~\eqref{eq:occupation_final} is
\begin{equation}
\widetilde n_0(\kappa)
=
\frac{1}{\pi}
\left[
\arctan\frac{\mu-\epsilon_\kappa}{\Gamma}
+
\frac{\pi}{2}
\right],
\label{eq:SM_n0_zeroT}
\end{equation}
and hence
\begin{equation}
\partial_\kappa \widetilde n_0(\kappa)
=
-\frac{\Gamma v_\kappa}
{\pi\left[
(\mu-\epsilon_\kappa)^2+\Gamma^2
\right]} .
\label{eq:SM_dn0_dk}
\end{equation}

Combining Eqs.~\eqref{eq:SM_delta_n_zeroT} and \eqref{eq:SM_dn0_dk}, we obtain the exact first-order relation
\begin{equation}
\delta \widetilde n_{\Gamma}(\kappa)
=
-
\frac{E\Gamma}
{(\mu-\epsilon_\kappa)^2+\Gamma^2}
\,
\partial_\kappa \widetilde n_0(\kappa).
\label{eq:SM_delta_n_exact_relation}
\end{equation}
Equivalently,
\begin{equation}
\widetilde n_{\Gamma}(\kappa)
=
\widetilde n_0(\kappa)
-
\Delta\kappa_{\Gamma}(\kappa)
\partial_\kappa \widetilde n_0(\kappa),
\end{equation}
with the momentum-dependent effective displacement
\begin{equation}
\Delta\kappa_{\Gamma}(\kappa)
=
\frac{E\Gamma}
{(\mu-\epsilon_\kappa)^2+\Gamma^2}.
\label{eq:SM_effective_shift_Gamma}
\end{equation}
The displacement is largest at the Fermi surface,
\begin{equation}
\Delta\kappa_{\Gamma}=\frac{E}{\Gamma}
\qquad
(\epsilon_\kappa=\mu),
\end{equation}
and rapidly decreases away from it. A closely related weak-field analysis for an electric-field-driven
tight-binding lattice coupled to fermion reservoirs was given in
Ref.~\cite{Han2013}. In that work, the field-induced displacement of the
distribution near the Fermi surface reduces to the estimate
$\Delta k \sim E/\Gamma$.

This result should be contrasted with the constant relaxation-time approximation. In that case, the distribution is assumed to be rigidly shifted by
\begin{equation}
\Delta\kappa_{\tau_B}=E\tau_B ,
\end{equation}
so that
\begin{equation}
\widetilde n_{\tau_B}(\kappa)
\simeq
\widetilde n_0(\kappa)
-
E\tau_B\,
\partial_\kappa \widetilde n_0(\kappa).
\label{eq:SM_relaxation_time_shift}
\end{equation}
Using the convention \(\tau_B=1/(2\Gamma)\), this gives a constant shift
\begin{equation}
\Delta\kappa_{\tau_B}=\frac{E}{2\Gamma}.
\label{eq:SM_Boltzmann_shift}
\end{equation}
The difference between the two approximations can be quantified by comparing the exact weak-field correction in Eq.~\eqref{eq:SM_delta_n_exact_relation} with the rigid-shift result in Eq.~\eqref{eq:SM_relaxation_time_shift}. For \(\tau_B=1/(2\Gamma)\), their ratio is
\begin{equation}
\frac{\Delta \kappa_{\Gamma}}
{\Delta \kappa_{\tau_B}}
=
\frac{2\Gamma^2}
{(\mu-\epsilon_\kappa)^2+\Gamma^2}.
\label{eq:SM_ratio_shift}
\end{equation}
This comparison also clarifies an important point. The constant-$\Gamma$
reservoir model does not in general generate a rigidly shifted Fermi distribution.
The constant-$\Gamma$ reservoir result is twice as large as the matched relaxation-time result at the Fermi surface, but is strongly suppressed away from the Fermi surface. In other words, within the weak-field expansion, the effective shift in the constant-$\Gamma$ reservoir model varies from its maximal value \(E/\Gamma\) at \(\epsilon_\kappa=\mu\) to nearly zero for \(|\epsilon_\kappa-\mu|\gg\Gamma\). A simple heuristic estimate of the effective shift is therefore obtained by averaging these two limiting values,
\begin{equation}
\frac{1}{2}
\left(
0+\frac{E}{\Gamma}
\right)
=
\frac{E}{2\Gamma},
\end{equation}
which coincides with the rigid shift in the constant-relaxation-time approximation. In this sense, although the two descriptions are not equivalent point by point in \(\kappa\), they lead to the same characteristic displacement scale in the small-\(\Gamma\) and weak-field limit.

This distinction between the two descriptions is unimportant in the strict weak-field limit if only the overall linear conductivity is considered.
However, it becomes important for estimating the field scale at which the nonperturbative regime starts.
In the constant-relaxation-time approximation, all occupied states are shifted by the same amount \(E\tau_B\).
In the constant-$\Gamma$ reservoir model, the relevant Fermi-surface displacement is instead \(E/\Gamma\).
Therefore, a criterion based on the Fermi-contour edge reaching a Berry-curvature hot spot~\cite{Sur2024FNE} 
predicts a smaller crossover field in the constant-$\Gamma$ reservoir model.
This explains why a rigid-shift Berry-curvature-dipole estimate can reproduce the weak-field trend but predict a different nonperturbative crossover.

We finally note that the constant-$\Gamma$ reservoir model is a single-particle dissipative description and does not include an electron-electron collision integral. Fast electron-electron scattering may drive the system closer to a locally thermal, or hot-electron, distribution, thereby reducing the detailed k-dependent displacement found in the constant-$\Gamma$ reservoir model. The present comparison should therefore be viewed as contrasting two limiting weak-field descriptions: a rigid constant-$\tau$ shift and a reservoir-induced constant-$\Gamma$ nonthermal distribution.

\section{Strong-field momentum distributions}

\begin{figure}[!tbp]
    \centering
    \includegraphics[width=\columnwidth]{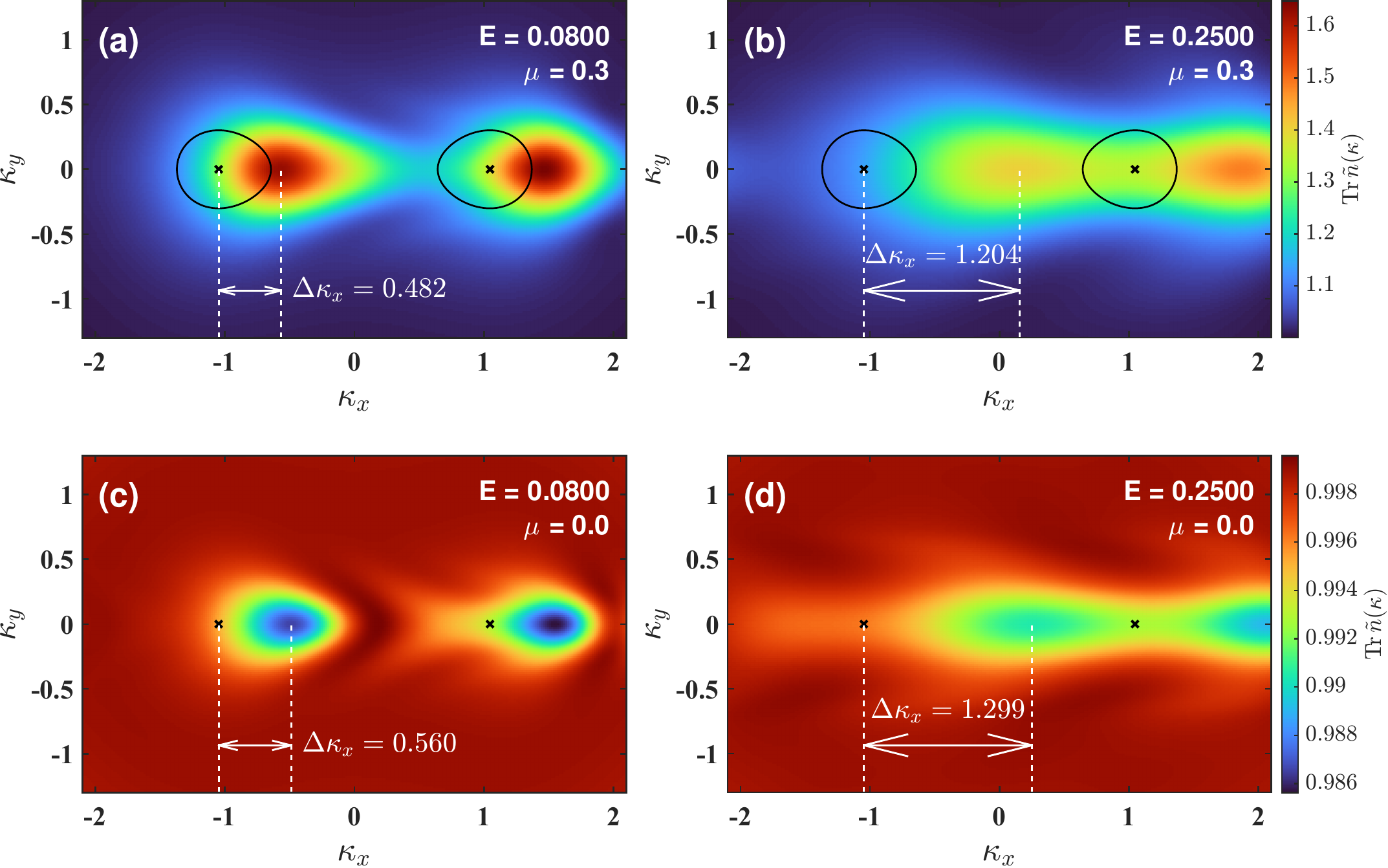}
\caption{
Momentum-resolved occupation distribution $\mathrm{Tr}\,\widetilde n(\boldsymbol{\kappa})$ in the strong-field regime. 
(a,b) Results for $\mu=0.3$ at $E=0.08$ and $E=0.25$, respectively. 
(c,d) Results for $\mu=0$ at $E=0.08$ and $E=0.25$, respectively. 
Black crosses mark the 
nodal points, and black contours in (a,b) indicate the equilibrium Fermi surface for $\mu=0.3$. 
White dashed lines and arrows denote the displacement $\Delta \kappa_x$ of the occupation maximum or minimum along the $\kappa_y\simeq0$ line relative to the left 
nodal point.
}
\label{fig:SM_strong_distribution}
\end{figure}

In the main text, we focused on the perturbative regime and on the onset of the nonperturbative regime in the momentum distribution $\widetilde n(\boldsymbol{\kappa})$, where the field-driven drift can still be estimated from the scale $E/\Gamma$.
Here we show additional momentum-space occupations at larger fields.
These results illustrate the far-from-perturbative regime, where the simple drift picture and the Berry-curvature-dipole interpretation are no longer sufficient.

Figure~\ref{fig:SM_strong_distribution} shows the momentum distributions for two representative chemical potentials, $\mu=0.3$ and $\mu=0$, at large fields $E=0.08$ and $E=0.25$. The imaginary part of the self-energy for the bath is $\Gamma=0.1$.
For both chemical potentials, increasing the field leads to a more broadly redistributed occupation in momentum space.
The maxima and minima associated with the nodal-point structures are displaced farther away from the nodal points, but the displacement no longer follows the simple estimate $E/\Gamma$.
In the panels we indicate the distance between the left
nodal point and the nearby left peak.
In the strong-field regime this distance is substantially smaller than $E/\Gamma$; for example, at $E=0.25$ it is only about half of the naive drift estimate.
Moreover, increasing the field further does not lead to a proportional increase of the displacement.
This indicates that the momentum-space drift saturates and that the simple drift picture breaks down in the far-from-perturbative regime.

Additional features also emerge at large fields.
First, the drift distances of the left and right structures are not identical.
This reflects the fact that the band structure on the two sides of the 
nodal points is not symmetric, so the field-induced redistribution depends on the local dispersion rather than only on the ratio $E/\Gamma$.
Second, the amplitudes of the left and right extrema become unequal, especially at $E=0.25$, where the left feature is significantly weaker than the right one.
These asymmetries show that the strong-field distribution cannot be understood in terms of a rigidly shifted weak-field pattern.

The semimetallic case $\mu=0$ displays an additional complication.
At $E=0.08$, the oscillatory tail of the hole-like structure generated near the left 
nodal point extends toward the structure associated with the right 
nodal point.
Thus the nonequilibrium structures generated near the two 
nodal  points begin to overlap and influence each other.
This overlap makes the field dependence of the Hall response more complex and contributes to the nonmonotonic, strongly nonperturbative behavior discussed in the main text.

\section{Robustness against weaker reservoir damping}

In the main text, we mainly use the damping rate $\Gamma=0.1$.
This value is convenient numerically, since it allows us to compute the nonequilibrium steady state within a moderate time and on a manageable momentum grid.
In clean metallic or semimetallic samples, however, the effective momentum scattering rate can be substantially smaller.
For example, terahertz measurements on the 
Dirac semimetal Cd$_3$As$_2$ reported a momentum scattering time of about \(157\) fs~\cite{Dai2021Cd3As2THz}.
To compare this value with the dimensionless damping in the lattice model, we estimate
\begin{equation}
\frac{\Gamma}{\hbar v_F/a}
\sim
\frac{a}{v_F\tau},
\end{equation}
up to a convention-dependent factor of order unity.
Using \(v_F\sim 1.5\times 10^6\,{\rm m/s}\)~\cite{neupane2014observation} and \(a\simeq 12.6\,\text{\AA}\)~\cite{mosca2017electronic}, this gives
\begin{equation}
\frac{a}{v_F\tau}
\sim 10^{-2}.
\end{equation}
In our lattice model, the lattice constant and the Dirac velocity are both of order unity, so that \(\Gamma/(\hbar v_F/a)\) is of the same order as the damping parameter \(\Gamma\).
This estimate indicates that clean Cd$_3$As$_2$ corresponds to a rather weak-damping regime of the lattice model.
It is therefore useful to check whether the nonperturbative features discussed in the main text remain visible at weaker damping.

\begin{figure}[!tbp]
    \centering
    \includegraphics[width=\columnwidth]{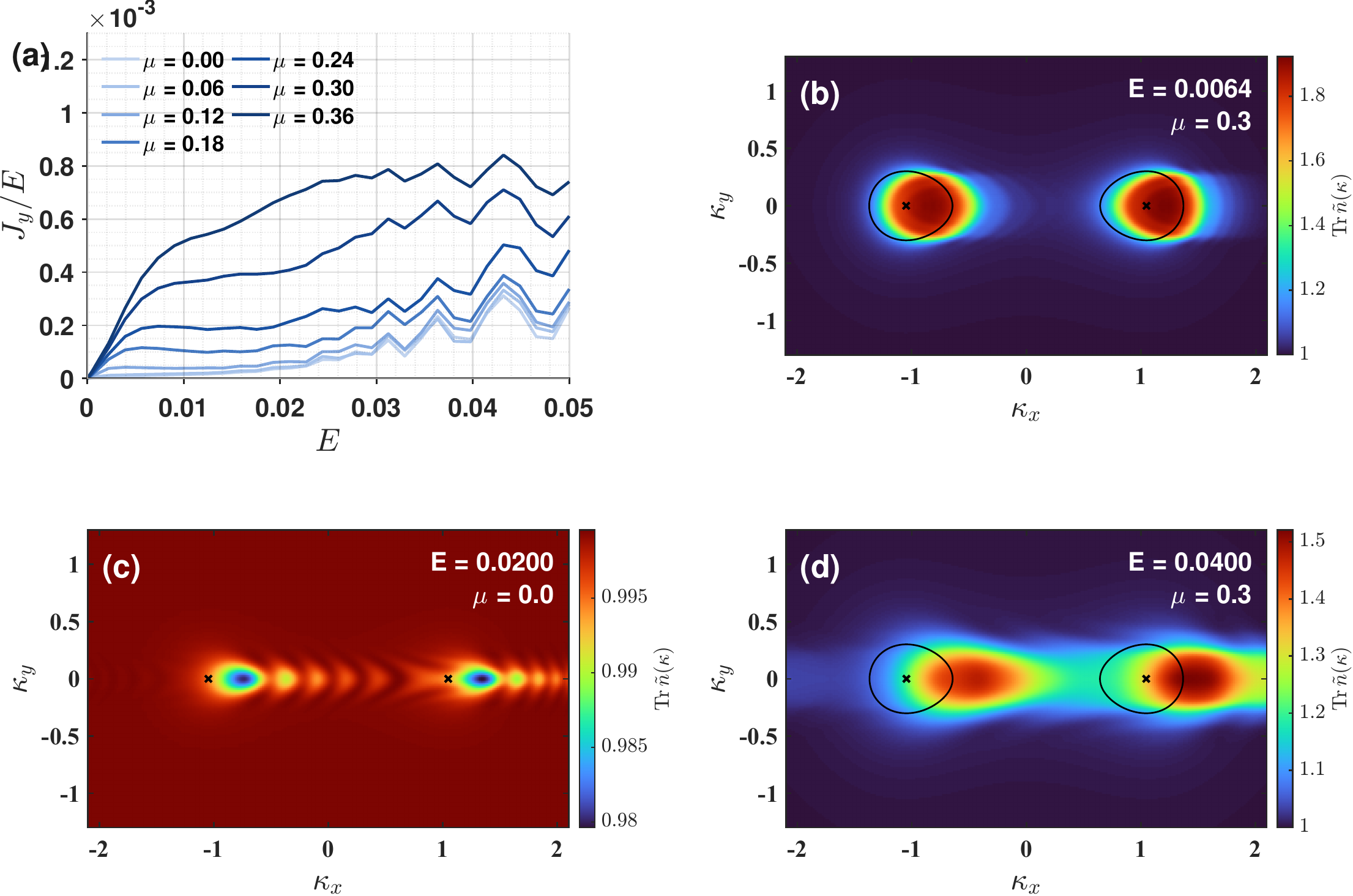}
\caption{
Results for weaker reservoir damping, $\Gamma=0.02$.
(a) Field-normalized transverse current $J_y/E$ as a function of the driving field $E$ for several chemical potentials $\mu$.
(b)--(d) Momentum-space occupation maps $\mathrm{Tr}\,\widetilde n(\boldsymbol{\kappa})$ at representative fields.
Panels (b) and (d) show results for $\mu=0.30$, with the upper-band Fermi contour and the 
nodal  points indicated.
Panel (c) shows the result for the case $\mu=0.00$ at $E=0.02$, with the
nodal  points marked.
}
\label{fig:SM_Gamma002}
\end{figure}

Figure~\ref{fig:SM_Gamma002}(a) shows the field dependence of $J_y/E$ for $\Gamma=0.02$ with different chemical potentials, corresponding to the results of Fig.~1(b) in the main text.
For nonzero chemical potentials, reducing $\Gamma$ shifts the drift-induced crossover to smaller fields, consistently with the characteristic displacement
$E/\Gamma$. In the nonperturbative regime, the oscillatory behavior of the Hall response is more pronounced than for $\Gamma=0.1$.
We attribute these oscillations in the Hall response to
Landau--Zener--St\"uckelberg interference
\cite{SHEVCHENKO20101,Lim2015}. 

For a fixed transverse momentum,
passage through a near-nodal region acts as a Landau--Zener event,
creating coherent amplitudes in the lower and upper bands. These two
components subsequently propagate along the field-driven momentum
trajectory and accumulate a relative dynamical phase. Neglecting the
Stokes phase associated with the individual Landau--Zener events and
retaining only the dynamical contribution, this phase can be written as
\begin{equation}
\Phi_{\mathrm{dyn}}(E)
=
\int_{t_1}^{t_2}
dt\,
\Delta\varepsilon[\boldsymbol{\kappa}(t)]
=
-\frac{1}{E}
\int_{\kappa_{1}}^{\kappa_{2}}
d\kappa_x\,
\Delta\varepsilon(\kappa_x,\kappa_y),
\label{eq:LZS_dynamic_phase}
\end{equation}
where $\Delta\varepsilon$ is the difference between the two band
energies and we have used $\dot{\kappa}_x=-E$. For interference between
a fixed pair of nodal points at fixed $\kappa_y$, the integration limits
and the momentum-space path are fixed. We therefore define
\begin{equation}
\mathcal{I}
=
\left|
\int_{\kappa_1}^{\kappa_2}
d\kappa_x\,
\Delta\varepsilon(\kappa_x,\kappa_y)
\right|,
\end{equation}
which is independent of $E$, so that
$|\Phi_{\mathrm{dyn}}(E)|=\mathcal{I}/E$. If the coherent propagation
spans a sufficient distance, the trajectory can encounter another nodal
region or return to the initial one owing to Brillouin-zone periodicity.
The upper- and lower-band amplitudes are then mixed again and interfere,
producing oscillations in the Hall response.

\begin{figure}[!tbp]
    \centering
    \includegraphics[width=\columnwidth]{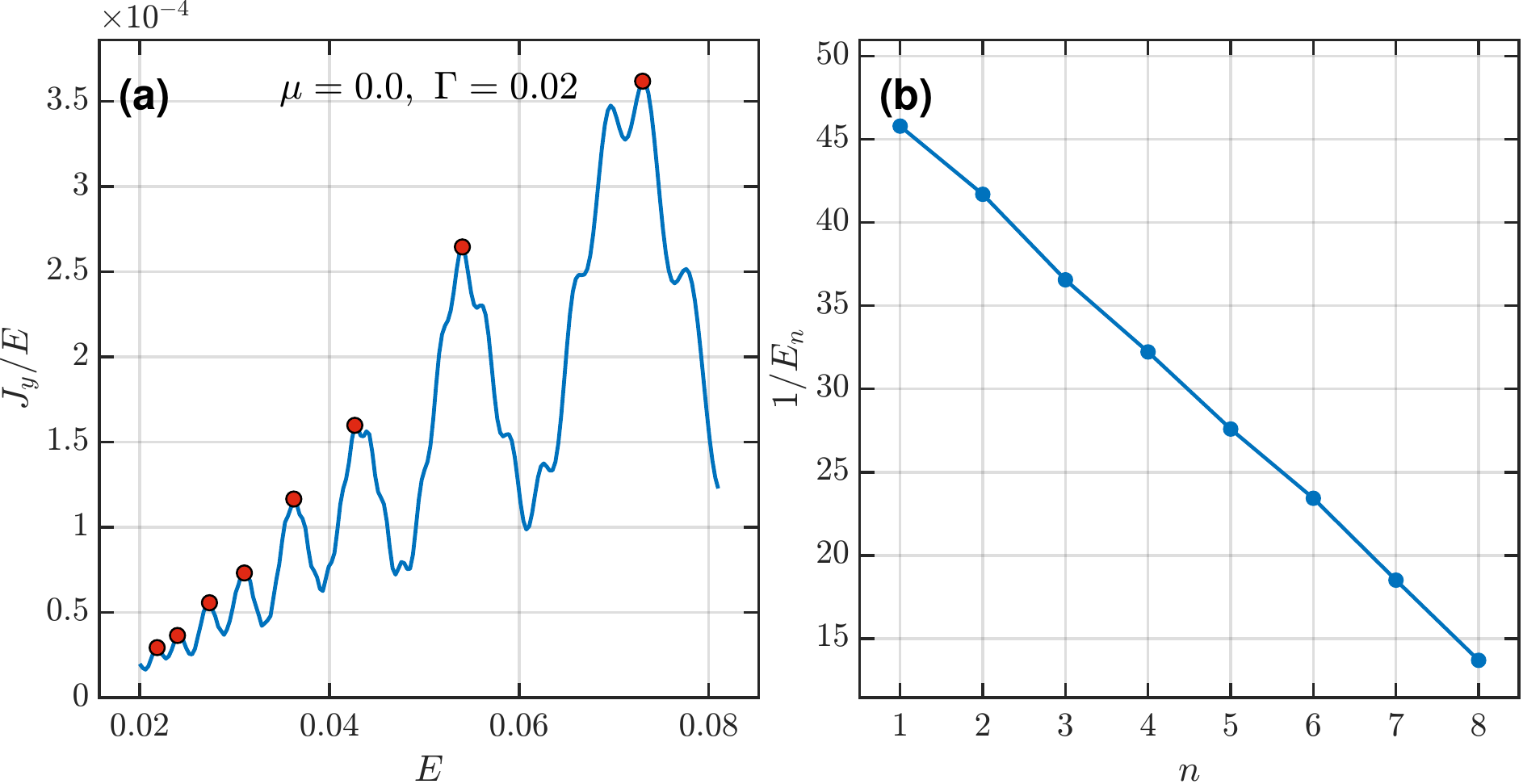}
    \caption{
    Field dependence of the oscillatory Hall response for
    $\mu=0$ and $\Gamma=0.02$.
    (a) Hall response $J_y/E$ as a function of the
    electric field. The symbols mark the successive peaks at fields
    $E_n$ used in the peak-position analysis.
    (b) Reciprocal peak field $1/E_n$ as a function of the peak index $n$.
    }
    \label{fig:SM_peak_spacing}
\end{figure}

The larger damping rate $\Gamma$ used in the main text limits the coherent
momentum-space propagation to
$\Delta\kappa\sim E/\Gamma$. Reducing $\Gamma$ 
allows the two coherent components to interfere at a
smaller electric field, i.e., it lowers the field scale at which
the interference oscillations become visible, while the peak positions
are governed by the accumulated interference phase discussed above.

To test this prediction, we denote the principal maxima of $J_y/E$
for $\mu=0$ and $\Gamma=0.02$ by $E_n$ in
Fig.~\ref{fig:SM_peak_spacing}(a). Within the dynamical-phase
approximation, constructive-interference maxima satisfy
\begin{equation}
|\Phi_{\mathrm{dyn}}(E_m)|=2\pi m,
\end{equation}
where $m$ is an integer interference order. Since
$|\Phi_{\mathrm{dyn}}(E)|=\mathcal{I}/E$, the corresponding peak fields
obey
\begin{equation}
\frac{1}{E_m}
=
\frac{2\pi m}{\mathcal{I}}.
\end{equation}
The peaks in Fig.~\ref{fig:SM_peak_spacing}(a) are indexed by increasing
electric field, so their interference order decreases successively and
may be written as $m=m_0-n$. Therefore,
\begin{equation}
\frac{1}{E_n}
=
B-An,
\qquad
B=\frac{2\pi m_0}{\mathcal{I}},
\qquad
A=\frac{2\pi}{\mathcal{I}}>0.
\label{eq:LZS_peak_positions}
\end{equation}
As shown in Fig.~\ref{fig:SM_peak_spacing}(b), the calculated values of
$1/E_n$ decrease approximately linearly with $n$. This agreement
provides quantitative support for identifying the field-dependent
oscillations as Landau--Zener--St\"uckelberg interference.

Figures~\ref{fig:SM_Gamma002}(b)--\ref{fig:SM_Gamma002}(d) show the corresponding momentum-space occupations $\mathrm{Tr}\,\widetilde n(\boldsymbol{\kappa})$ for representative cases, analogous to Fig.~2 in the main text and Fig.~\ref{fig:SM_strong_distribution}.
Figure~\ref{fig:SM_Gamma002}(b) shows the result for $\mu=0.3$ and $E=0.0064$.
According to the simple drift estimate $\Delta \kappa_x\sim E/\Gamma$, this field corresponds to a displacement comparable to the distance between the Fermi contour edge and the 
nodal point.
One might therefore expect, by analogy with Fig.~2, that the edge of the high-occupation region is shifted to the
nodal  point.
The actual distribution in Fig.~\ref{fig:SM_Gamma002}(b), however, is more subtle.
Although a narrow line-like structure appears near the 
nodal  point, the boundary of the high-occupation region still encloses it, rather than being simply displaced to it.
This behavior is a consequence of the constant-$\Gamma$ reservoir model.
As shown in Eq.~\eqref{eq:SM_effective_shift_Gamma}, the effective displacement is energy dependent: it is maximal at the Fermi surface, where it equals $E/\Gamma$, and decreases rapidly away from the chemical potential.
For smaller $\Gamma$, this decrease becomes sharper, so that essentially only the outermost layer of the broadened Fermi edge is shifted by the full amount $E/\Gamma$.
The total occupation edge therefore does not behave as a rigidly shifted contour and need not coincide with the 
nodal  point. Nevertheless, Fig.~\ref{fig:SM_Gamma002}(a) shows that the same field scale still coincides well with the turning point of the Hall response.
This confirms that the onset of the nonperturbative feature is controlled by the displacement of the Fermi-surface layer, rather than by the apparent boundary of the full momentum occupation.

Figure~\ref{fig:SM_Gamma002}(c) shows the semimetallic case
$\mu=0$ at $E=0.02$. Compared with the corresponding
$\Gamma=0.1$ distributions discussed above, the oscillatory structures
are considerably more pronounced. Finally, Fig.~\ref{fig:SM_Gamma002}(d) shows the case with $\mu=0.3$ at $E=0.04$, already in the deep nonperturbative regime.
The overall structure is consistent with the strong-field distributions shown in Fig.~\ref{fig:SM_strong_distribution}, but the momentum displacement has a smaller saturated drift.

Overall, these smaller-$\Gamma$ results show that the energy-dependent effective drift in the constant-$\Gamma$ reservoir model becomes more important for the detailed momentum-space structure when damping is weak.
Nevertheless, the main nonperturbative features identified in the main text remain robust, indicating that our conclusions are also relevant for cleaner materials with smaller scattering rates.

\section{Momentum-resolved density matrix in different bases}

In the main text, we visualize the nonequilibrium momentum distribution through the total occupation $\mathrm{Tr}\,\widetilde n(\boldsymbol{\kappa}) = \widetilde n_{11}(\boldsymbol{\kappa})+\widetilde n_{22}(\boldsymbol{\kappa})$, as shown in Fig.~2.
This quantity is basis independent and contains the combined contribution from the two orbital components, or from the upper and lower bands.
To gain more microscopic insight, we further decompose the momentum-resolved density matrix into its individual matrix elements in different bases.
This analysis is particularly useful for the semimetal case $\mu=0$, where the Hall response is governed by interband charge transfer rather than by a conventional Fermi-surface drift.

\begin{figure}[!tbp]
    \centering
    \includegraphics[width=\columnwidth]{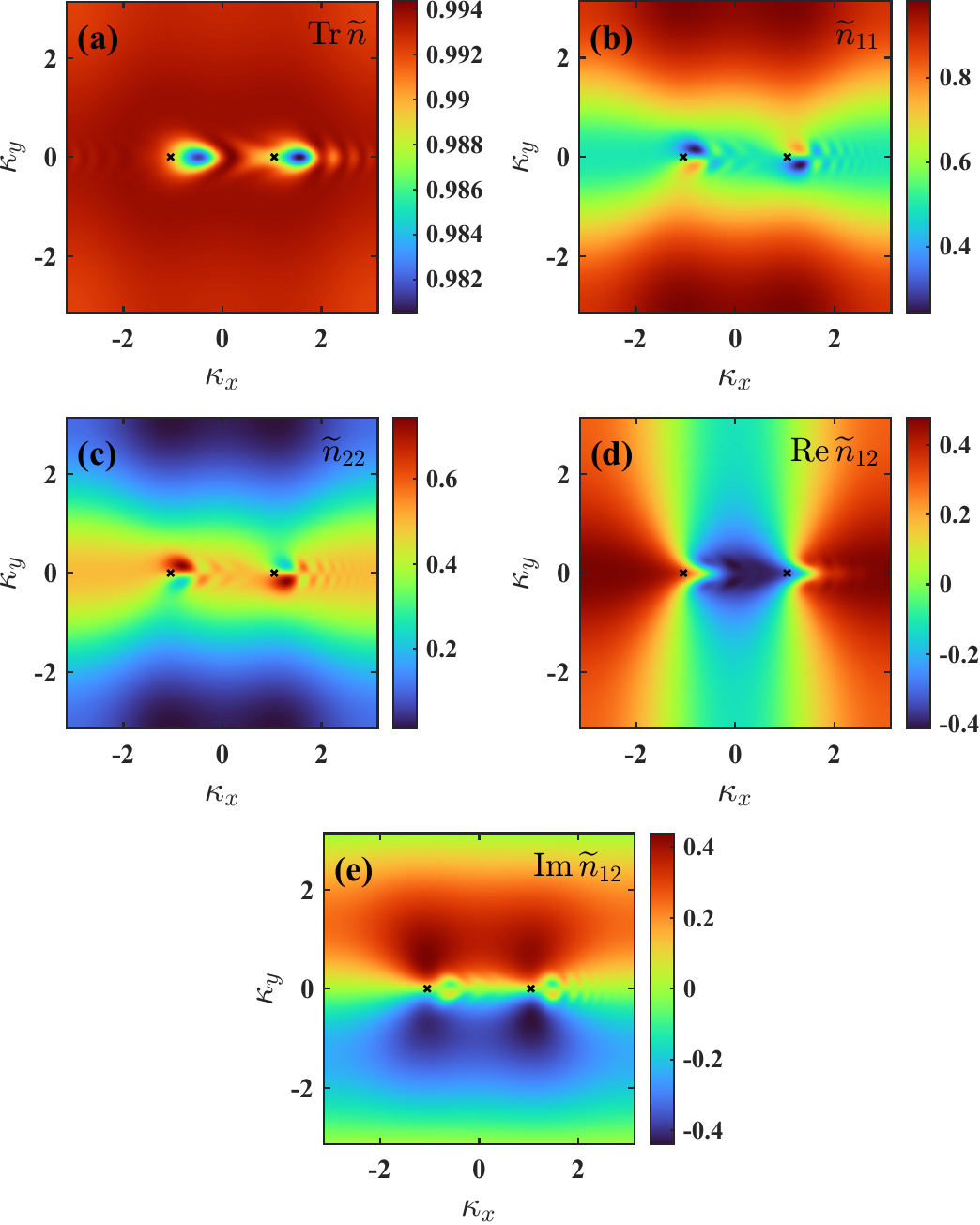}
\caption{
Momentum-resolved density matrix in the orbital basis for the case with  $\mu=0$ and $\Gamma=0.1$ at driving field $E=0.08$. The panels show the trace of the occupation matrix, $\mathrm{Tr}\,\widetilde n(\boldsymbol{\kappa}) = \widetilde n_{11}(\boldsymbol{\kappa})+\widetilde n_{22}(\boldsymbol{\kappa})$~(a), together with the orbital-basis matrix elements $\widetilde n_{11}$~(b), $\widetilde n_{22}$~(c), $\mathrm{Re}\,\widetilde n_{12}$~(d), and $\mathrm{Im}\,\widetilde n_{12}$~(e). Black crosses mark the positions of the two 
nodal points. 
}
\label{fig:SM_density_orbital}
\end{figure}

We first analyze the density matrix in the original orbital basis, i.e., the basis in which the Hamiltonian in the main text is written.
As a representative example, we choose $\mu=0$, $E=0.08$, and $\Gamma=0.1$.
The corresponding momentum-resolved matrix elements are shown in Fig.~\ref{fig:SM_density_orbital}.
Since the density matrix is Hermitian, we only show the real parts of the two diagonal elements, $\mathrm{Re}\,\widetilde n_{11}(\boldsymbol{\kappa})$ and $\mathrm{Re}\,\widetilde n_{22}(\boldsymbol{\kappa})$, together with the real and imaginary parts of the off-diagonal element, $\mathrm{Re}\,\widetilde n_{12}(\boldsymbol{\kappa})$ and $\mathrm{Im}\,\widetilde n_{12}(\boldsymbol{\kappa})$.
For comparison, the total occupation $\mathrm{Tr}\,\widetilde n(\boldsymbol{\kappa})$ is also displayed.

The two diagonal orbital components have distinct momentum-space distributions.
The first orbital mainly contributes to the occupation near $\kappa_y=\pi$, whereas the second orbital mainly contributes near $\kappa_y=0$.
Under the applied electric field, pronounced nonequilibrium structures are generated around the nodal points.
These structures are visible not only in the diagonal components but also in both the real and imaginary parts of the off-diagonal density matrix. Another feature is that the total occupation $\mathrm{Tr}\,\widetilde n(\boldsymbol{\kappa})$ remains symmetric with respect to $\kappa_y=0$, while the individual orbital occupations $\widetilde n_{11}(\boldsymbol{\kappa})$ and $\widetilde n_{22}(\boldsymbol{\kappa})$ do not separately exhibit this symmetry.

\begin{figure}[!tbp]
    \centering
    \includegraphics[width=\columnwidth]{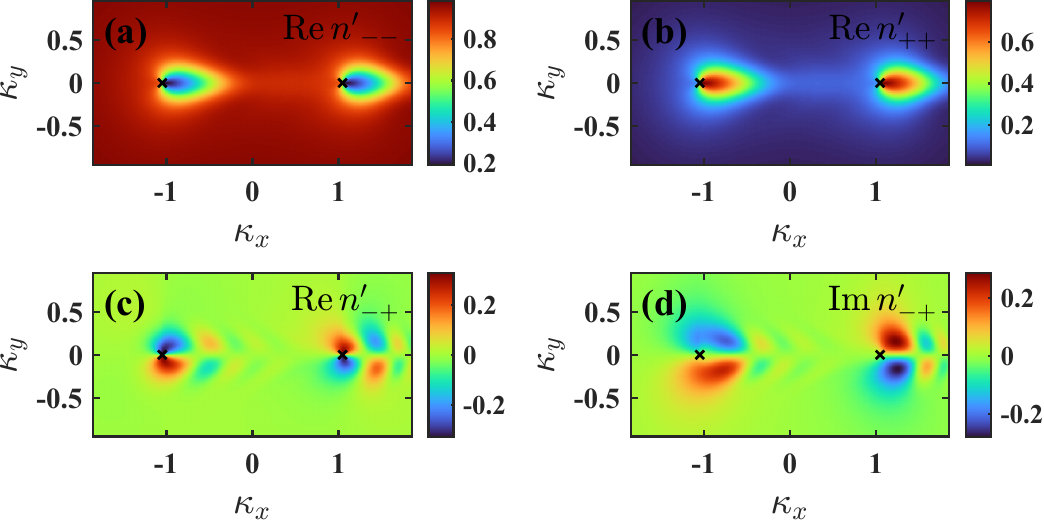}
\caption{
Momentum-resolved density matrix in the diagonalized basis for the model with  $\mu=0$ and $\Gamma=0.1$ at driving field $E=0.08$. The panels show the trace of the occupation matrix elements ${n'}_{--}$~(a), ${n'}_{++}$~(b), $\mathrm{Re}\,{n'}_{-+}$~(c), and $\mathrm{Im}\,{n'}_{-+}$~(d). Black crosses mark the positions of the two nodal points. 
}
\label{fig:SM_density_band}
\end{figure}

Although the orbital-basis decomposition already reveals the structure of the density matrix, it is not the most intuitive representation for discussing interband charge transfer.
It is therefore useful to transform the density matrix to the equilibrium band basis, i.e., the basis that diagonalizes the field-free Hamiltonian,
\begin{equation}
U^\dagger(\boldsymbol{\kappa}) H(\boldsymbol{\kappa}) U(\boldsymbol{\kappa})
=
\begin{pmatrix}
\varepsilon_-(\boldsymbol{\kappa}) & 0 \\
0 & \varepsilon_+(\boldsymbol{\kappa})
\end{pmatrix}.
\end{equation}
The density matrix in this basis is
\begin{equation}
n'(\boldsymbol{\kappa})
=
U^\dagger(\boldsymbol{\kappa}) \widetilde n(\boldsymbol{\kappa}) U(\boldsymbol{\kappa}) .
\label{diag_trnas}
\end{equation}
In equilibrium, \(n'(\boldsymbol{\kappa})\) is diagonal in the equilibrium band basis, where the \(+\) and \(-\) states denote the eigenstates of the field-free Hamiltonian. In other words, the off-diagonal matrix elements vanish at \(E=0\) only after transforming to this \(+/-\) basis. This makes the field-induced interband coherence directly visible. For the semimetallic case \(\mu=0\) considered here, the equilibrium state is close to a filled lower band and an empty upper band in the small-\(\Gamma\) limit, up to spectral-broadening effects near the nodal points. The field-induced changes of the diagonal populations and off-diagonal coherences are therefore especially transparent in the equilibrium band basis.

In performing the transformation in Eq.~\eqref{diag_trnas}, one has to fix the gauge of the eigenvectors at different momenta.
A direct diagonalization at each $\boldsymbol{\kappa}$ point leaves an arbitrary momentum-dependent phase, which can make the off-diagonal element $n'_{+-}(\boldsymbol{\kappa})$ appear artificially discontinuous.
We therefore choose a smooth gauge by requiring the overlap between eigenvectors at neighboring $\boldsymbol{\kappa}$ points to be real and positive.
This procedure removes rapid phase jumps in momentum space, although an overall phase freedom remains.

The resulting band-basis density matrix is shown in Fig.~\ref{fig:SM_density_band}.
Since the trace is invariant under the unitary transformation, $\mathrm{Tr}\,n'(\boldsymbol{\kappa})=\mathrm{Tr}\,\widetilde n(\boldsymbol{\kappa})$, it is not plotted again.
Figures~\ref{fig:SM_density_band}(a) and \ref{fig:SM_density_band}(b) show the diagonal elements of $n'(\boldsymbol{\kappa})$, corresponding to the lower- and upper-band occupations.
Both are symmetric with respect to $\kappa_y=0$.
Unlike the hole-like structure in the total occupation, whose center is shifted away from the nodal point, the strongest changes in the individual band occupations occur near the nodal points themselves.
This is expected, because the band gap vanishes there and interband charge transfer is most efficient.
The shifted hole-like structure in $\mathrm{Tr}\,\widetilde n(\boldsymbol{\kappa})$ therefore results from the imbalance between the field-induced lower- and upper-band population changes.

Figures~\ref{fig:SM_density_band}(c) and \ref{fig:SM_density_band}(d) show the real and imaginary parts of the off-diagonal element $n'_{+-}(\boldsymbol{\kappa})$.
These components measure the interband coherence in the equilibrium band basis.
They are concentrated near the nodal points, are not symmetric with respect to $\kappa_y=0$, and exhibit oscillatory structures.
In the following, we will show that the Hall current 
originates entirely from the interband contribution.

In the band basis, the transverse current is
\begin{equation}
J_y
=
\int_{\rm BZ}\frac{d^2 \kappa}{(2\pi)^2}
\,
\mathrm{Tr}\left[
 v'_y(\boldsymbol{\kappa}) n'(\boldsymbol{\kappa})
\right],
\end{equation}
where
\begin{equation}
v'_y(\boldsymbol{\kappa})
=
U^\dagger(\boldsymbol{\kappa})\,\partial_{\kappa_y}H(\boldsymbol{\kappa})\,U(\boldsymbol{\kappa}) .
\end{equation}
Writing the band indices as $-$ and $+$, this becomes
\begin{equation}
J_y
=
\int_{\rm BZ}\frac{d^2 \kappa}{(2\pi)^2}
\left[
v'^y_{--} n'_{--}
+
v'^y_{++} n'_{++}
+
2\,\mathrm{Re}\left(
v'^y_{+-} n'_{-+}
\right)
\right].
\label{eq:Jy_band_decomposition}
\end{equation}
The diagonal velocity matrix elements are simply the band velocities,
\begin{equation}
v'^y_{\pm\pm}(\boldsymbol{\kappa})
=
\partial_{\kappa_y}\varepsilon_\pm(\boldsymbol{\kappa})
=
\pm\,\partial_{\kappa_y}\varepsilon(\boldsymbol{\kappa}),
\end{equation}
with $\varepsilon(\boldsymbol{\kappa})=|\mathbf{d}(\boldsymbol{\kappa})|$.
For the model used in the main text,
\begin{equation}
\mathbf{d}(\boldsymbol{\kappa})
=
\bigl(
(\cos \kappa_x-D_x),\,
\sin \kappa_y,\,
(\cos \kappa_y-D_y)
\bigr),
\end{equation}
and hence
\begin{equation}
\partial_{\kappa_y}\varepsilon(\boldsymbol{\kappa})
=
\frac{
D_y\sin \kappa_y
}{
\varepsilon(\boldsymbol{\kappa})
},
\end{equation}
which is odd under $\kappa_y\to -\kappa_y$.
Since the diagonal occupations $ n'_{--}$ and $ n'_{++}$ are even in $\kappa_y$, their contributions to $J_y$ vanish after integration over the Brillouin zone.
Therefore, for the present symmetry setting, the transverse Hall current is carried by the off-diagonal term in Eq.~\eqref{eq:Jy_band_decomposition}.

\begin{figure}[!tbp]
    \centering
    \includegraphics[width=\columnwidth]{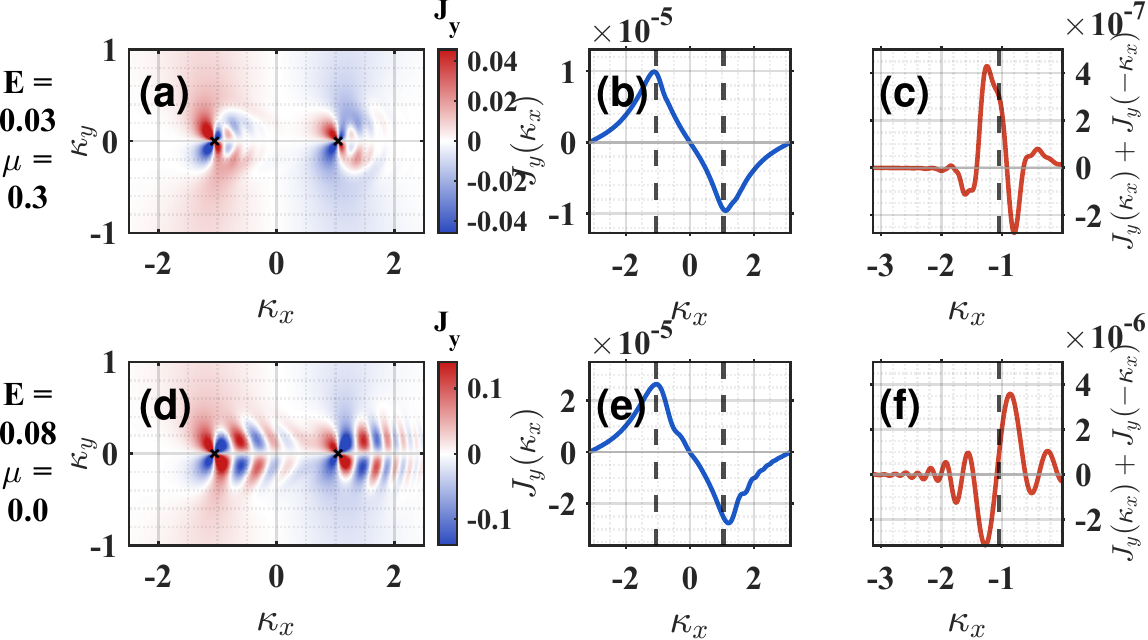}
\caption{
Momentum-resolved off-diagonal contribution to the Hall current in the field-free band basis.
The upper and lower rows correspond to $(E,\mu)=(0.03,0.3)$ and $(0.08,0.0)$, respectively.
Panels (a,d) show $J_y(\kappa_x,\kappa_y)$, panels (b,e) show the $\kappa_y$-integrated $J_y(\kappa_x)$, and panels (c,f) show $J_y(\kappa_x)+J_y(-\kappa_x)$. Vertical dotted lines mark the positions of nodal points.
}
\label{fig:SM_jy_distribution}
\end{figure}

To further inspect how the off-diagonal density matrix generates the Hall current, we plot the momentum-resolved contribution 
of the last term in
Eq.~\eqref{eq:Jy_band_decomposition}.
Its Brillouin-zone integral gives the total transverse current.
We consider two representative parameter sets, $(E,\mu)=(0.03,0.3)$ and $(E,\mu)=(0.08,0)$, shown in Figs.~\ref{fig:SM_jy_distribution}(a)--(c) and Figs.~\ref{fig:SM_jy_distribution}(d)--(f), respectively. The reservoir damping parameter is $\Gamma=0.1$.

For each case, we first show the two-dimensional distribution $J_y(\boldsymbol{\kappa})$.
We then integrate over $\kappa_y$ to get ${J}_y(\kappa_x)$,
and finally fold the positive and negative $\kappa_x$ axes by plotting ${J}_y(\kappa_x)+{J}_y(-\kappa_x)$.
This folded quantity highlights the residual imbalance between opposite momenta that survives the cancellation in the Brillouin-zone integral.

Although Eq.~\eqref{eq:Jy_band_decomposition} shows that the Hall current is carried by the off-diagonal term, Fig.~\ref{fig:SM_jy_distribution} also demonstrates why a simple momentum-space picture analogous to the Berry-curvature-dipole interpretation is difficult to construct in the full Hall current calculation.
The two-dimensional distributions in Figs.~\ref{fig:SM_jy_distribution}(a) and \ref{fig:SM_jy_distribution}(d) are highly oscillatory.
The net contribution can be attributed to the asymmetry after the $\kappa_y$ integration in Figs.~\ref{fig:SM_jy_distribution}(b) and \ref{fig:SM_jy_distribution}(e). But oscillations emerge again after folding the positive and negative $\kappa_x$ axes in Figs.~\ref{fig:SM_jy_distribution}(c) and \ref{fig:SM_jy_distribution}(f).
The dominant contributions are concentrated around the nodal points, but the net Hall current results from a delicate cancellation of oscillatory structures, making it difficult to assign the current to a single localized feature.
The semimetallic case $\mu=0$ shows a cleaner oscillatory pattern than the finite-density case $\mu=0.3$, where additional structures arise from the existing Fermi surface. Owing to the momentum-space structure discussed above, the Brillouin-zone-integrated Hall current converges on a relatively modest momentum grid. For the finite-field results considered here, a $100\times100$ grid is sufficient, substantially smaller than the grids required in our perturbative Kubo evaluation~\cite{Michishita2022PRB}. This favorable convergence constitutes a practical numerical advantage of the NEGF approach.
\section{Field and damping dependence of the interband mechanism}

In the previous section, the density matrix in the equilibrium band basis revealed the interband character of the response in the semimetallic case $\mu=0$.
This interband contribution cannot be reliably captured by a quasiparticle anomalous-velocity formula based only on Berry curvature, and therefore lacks a simple Fermi-surface-shift interpretation.
Here, we analyze how it depends on the electric field and on the damping rate $\Gamma$.
This provides a complementary view of the interband charge-transfer process and clarifies the different behavior of the longitudinal and transverse currents.

\begin{figure}[!tbp]
    \centering
    \includegraphics[width=\columnwidth]{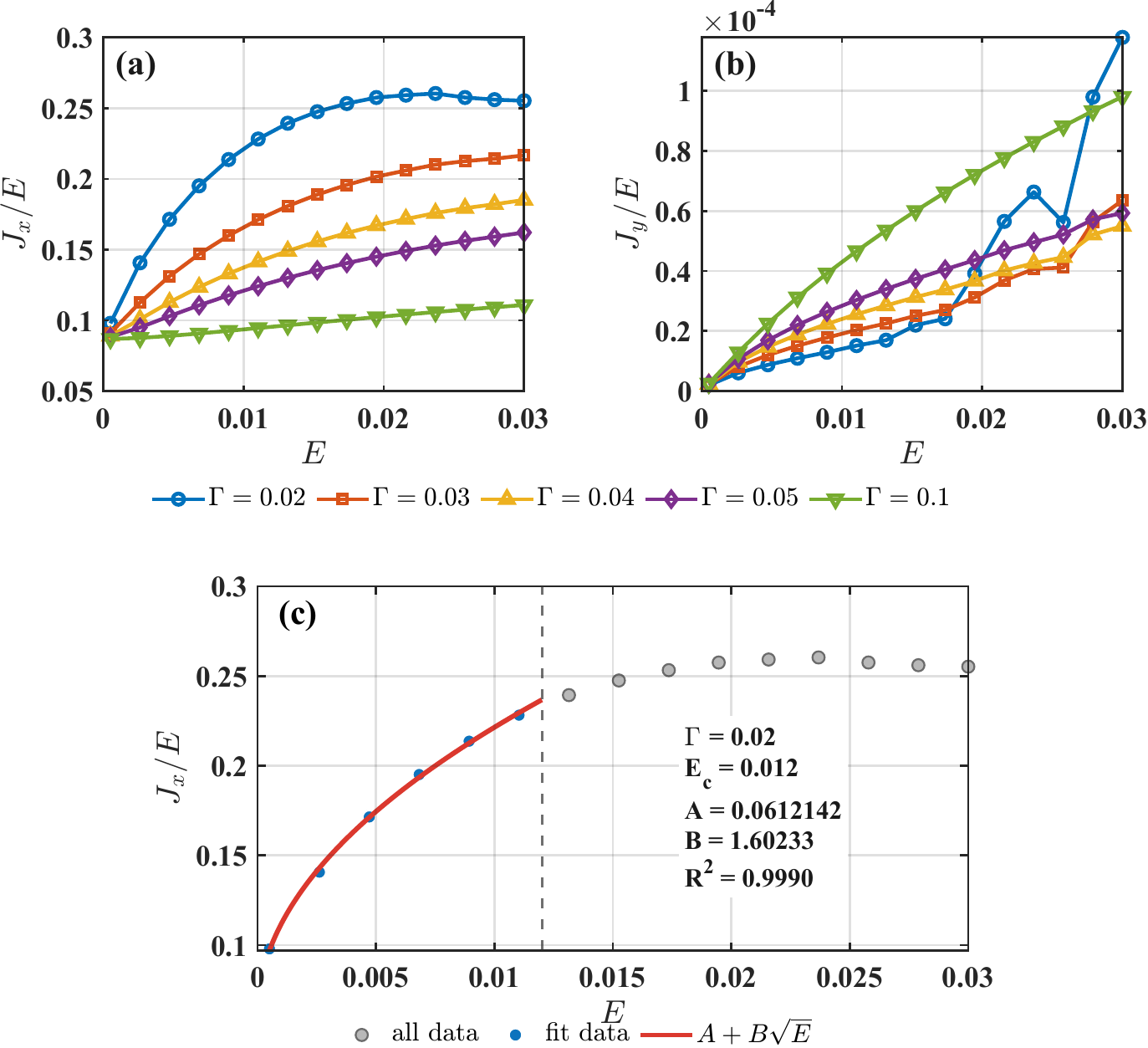}
\caption{
Field and damping dependence of the current response for $\mu=0$.
(a) Longitudinal response $J_x/E$ and (b) transverse response $J_y/E$
for different reservoir damping parameters $\Gamma$.
(c) Weak-field fit of the longitudinal response for $\Gamma=0.02$
to $J_x/E=A+B\sqrt{E}$ using data with $E<E_c=0.012$.
}
\label{fig:SM_mu0_Gamma}
\end{figure}

Figure~\ref{fig:SM_mu0_Gamma}(a) shows the longitudinal response $J_x/E$ for several values of $\Gamma$, while Fig.~\ref{fig:SM_mu0_Gamma}(b) shows the corresponding transverse response $J_y/E$.
Interestingly, the two quantities show opposite trends with damping.
The longitudinal conductivity decreases as $\Gamma$ is increased, whereas the transverse response becomes larger.
Their field dependences are also qualitatively different.
The longitudinal conductivity has a finite intercept in the weak-field limit and then increases nonlinearly, while $J_y/E$ has no intercept. For $\Gamma=0.1$, $J_y/E$ is approximately linear in $E$ over the low-field window, while smaller $\Gamma$ produce sublinear effective exponents over the same finite field range.

We first discuss the longitudinal response.
For the smallest damping rate considered here, $\Gamma=0.02$, the weak-field behavior of $J_x/E$ is well described by
\begin{equation}
\frac{J_x(E)}{E}
\simeq
A+B\sqrt{E},
\label{eq:sigma_x_sqrt_fit}
\end{equation}
as shown in Fig.~\ref{fig:SM_mu0_Gamma}(c).
The constant term $A$ represents the conductivity at the Dirac point in the weak-field limit.
This is analogous to the situation in graphene, where the dc response at the Dirac point contains an interband polarization contribution and cannot be interpreted simply as a Drude response from a Fermi surface~\cite{Carbotte2010,Dora2010}. The $\sqrt{E}$ correction can be understood as a signature of field-induced carrier production.
For an ideal two-dimensional Dirac cone, the Schwinger mechanism gives a field-induced carrier excitation $\Delta n(E)\propto E^{3/2}$,
as discussed for graphene in Refs.~\cite{Dora2010,Allor2008}.
These field-induced carriers then contribute to the longitudinal current in the presence of relaxation, which gives
\begin{equation}
\Delta J_x(E)\propto \frac{\Delta n(E)}{\Gamma}
\propto \frac{E^{3/2}}{\Gamma}.
\label{eq:carrier_current}
\end{equation}
Consequently, for fixed $\Gamma$, the corresponding correction to the longitudinal conductivity scales as $\sqrt{E}$, consistent with the fit in Fig.~\ref{fig:SM_mu0_Gamma}(c).
As $\Gamma$ is increased, the field-induced current is reduced by faster relaxation, which explains why the finite-field correction to $\sigma_x$ decreases with increasing damping. By contrast, the weak-field intercept is much less sensitive to $\Gamma$. This behavior is consistent with the interband contribution obtained from the Kubo formula at the Dirac point~\cite{Carbotte2010}.
Although Ref.~\cite{Dora2010} studies the real-time dynamics of isolated graphene rather than a dissipative steady state, the separation between the weak-field interband contribution and the field-induced carrier-production contribution provides a useful interpretation of our results.

\begin{figure}[!tbp]
    \centering
    \includegraphics[width=\columnwidth]{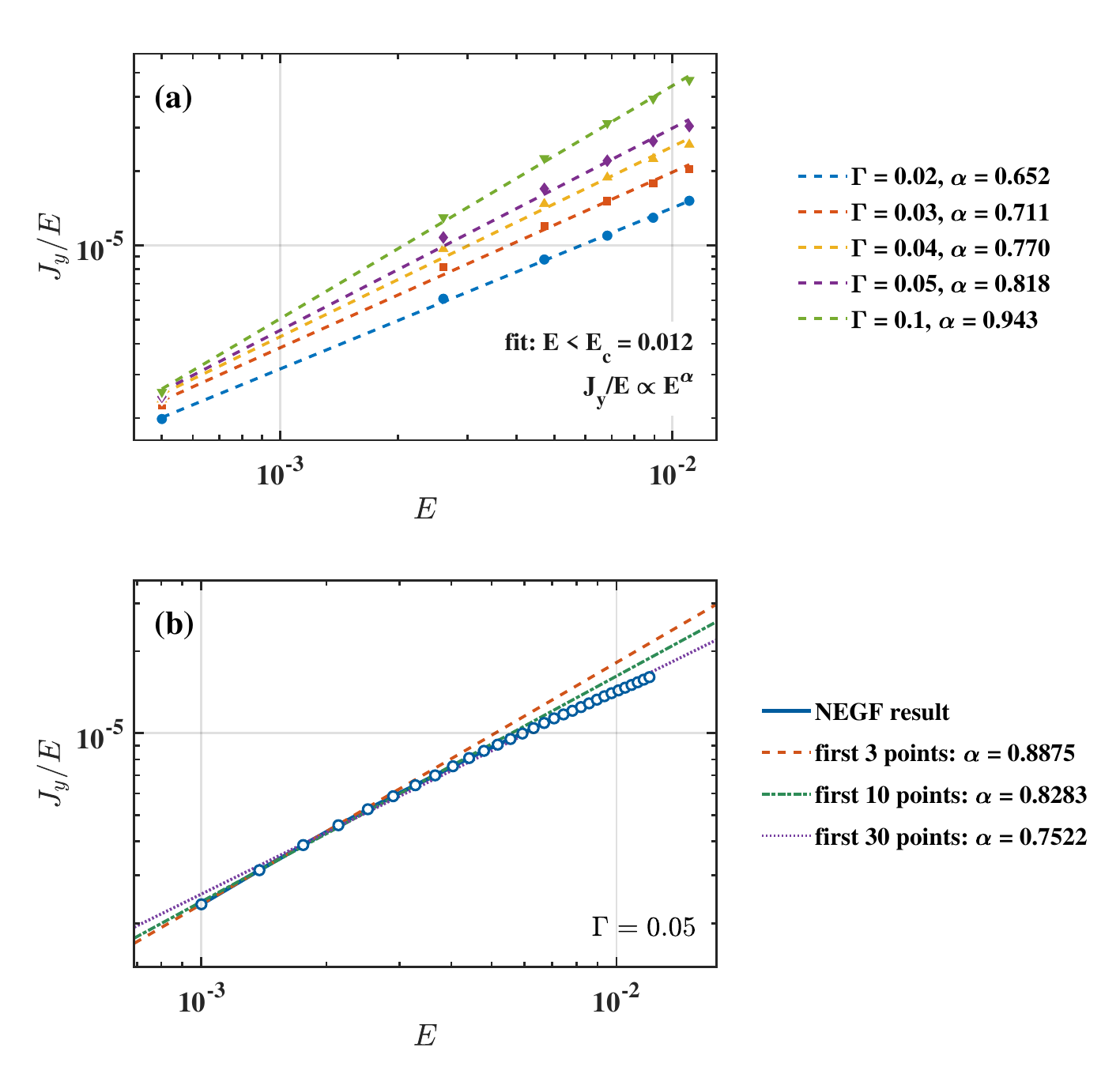}
\caption{
Weak-field scaling and fitting-window dependence of the transverse
response $J_y/E$ for $\mu=0$.
(a) Log-log plot for different reservoir damping parameters $\Gamma$.
Dashed lines show fits to $J_y/E\propto E^\alpha$ using data with
$E<E_c=0.012$; the fitted exponents are indicated in the legend.
(b) Fitting-window dependence for $\Gamma=0.05$ using more data points.
The circles show the NEGF result, and the dashed, dash-dotted, and
dotted lines show fits using the first 3, 10, and 30 data points from
the lowest electric field, with $\alpha=0.8875$, $0.8283$, and
$0.7522$, respectively.
}
\label{fig:SM_mu0_transverse}
\end{figure}

For the transverse response, Fig.~2(d) of the main text shows that
an estimate based solely on the change in the Berry-curvature-weighted
band occupations due to carrier production fails to reproduce the
full NEGF result. We therefore examine the field scaling
and damping dependence of the transverse response to further
characterize its behavior and gain insight into the underlying
mechanism.

Figure~\ref{fig:SM_mu0_transverse}(a) shows $J_y/E$ on a log-log
scale. For the main-text value $\Gamma=0.1$, the effective exponent
over the accessible low-field window is $\alpha=0.943$, which is
close to the perturbative value $\alpha=1$ corresponding to
$J_y\propto E^2$. For smaller damping rates, fits over the same
absolute field window yield smaller effective exponents.
To assess their sensitivity to the fitting window,
Fig.~\ref{fig:SM_mu0_transverse}(b) uses additional low-field data for
$\Gamma=0.05$. Fits to the first 30, 10, and 3 data points, ordered
from the lowest field upward, give
$\alpha=0.7522$, $0.8283$, and $0.8875$, respectively.
The effective exponent thus moves toward $\alpha=1$ as the fitting
window narrows.

The damping and fitting-window dependences are consistent with a
gradual crossover out of the perturbative regime, rather than distinct
asymptotic weak-field power laws. Since the field-induced redistribution
is controlled by $E/\Gamma$, reducing $\Gamma$ lowers the characteristic
field scale for this crossover, so the same absolute fitting window
samples a larger portion of the nonperturbative regime. In the present
semimetallic case, the crossover has no sharply defined critical field;
we therefore characterize it through effective exponents over finite
field windows. These fits are not intended to establish the
exact $E\to0$ asymptotic scaling.

Complementary to the field-scaling analysis, the weak-field response
$J_y/E$ increases approximately linearly with $\Gamma$, opposite to the
longitudinal conductivity. A possible microscopic origin is the generalized
Berry-curvature contribution $\sigma_{\mathrm{gBC}}$ identified in the
second-order multiband Green's function theory of
Ref.~\cite{Michishita2022PRB}. Whereas ordinary Berry curvature involves
first momentum derivatives of Bloch states, the generalized quantity
couples first and second derivatives through intermediate band states,
encoding interband information beyond a Berry-curvature-weighted occupation.
Finite-lifetime broadening enters the interband energy denominators,
giving a dissipation-induced geometric contribution whose reported leading
scaling is $\sigma_{\mathrm{gBC}}\propto\tau^{-1}\propto\Gamma$.
Its contribution to the field-normalized current therefore scales as
$J_y/E\propto\sigma_{\mathrm{gBC}}E$, consistent with the observed damping
trend. This differs from the conventional Berry-curvature-dipole term,
which scales as $\tau$ and relies on a Fermi-surface-shift picture
inapplicable at $\mu=0$.

Taken together, these results provide complementary perspectives
on the longitudinal and transverse responses in the interband regime.
The longitudinal response admits an interpretation in terms
of a weak-field interband contribution and field-induced carrier
production. For the transverse response, the weak-field damping
dependence is consistent with a generalized-Berry-curvature
contribution, providing a possible microscopic interpretation in the
perturbative regime. The field-scaling analysis further supports a
gradual crossover toward nonperturbative behavior.
\section{Effects from electron-phonon coupling and local interactions}

In the main text, we showed that electron-phonon coupling and local interactions can substantially modify the nonlinear Hall response.
Here we describe how these effects are incorporated into the steady-state NEGF calculation and clarify their physical origin.
In our framework, both effects enter through local self-energies.
These self-energies affect the Hall response not only through their imaginary parts, which describe additional damping, but also through their real parts, which renormalize the effective band structure and hence the Berry-curvature distribution.

In the calculation, we start from the converged nonequilibrium Green's function of the noninteracting system coupled to the fermionic bath.
This solution is then used as the initial input for a DMFT self-consistency loop.
At each iteration, the local Green's function is computed as
\begin{equation}
G_{\rm loc}(\omega)
=
\int_{\rm BZ}\frac{d^2 \kappa}{(2\pi)^2}
\widetilde G(\omega,\boldsymbol{\kappa}) ,
\end{equation}
and the electron-phonon or interaction self-energy is evaluated from this local Green's function.
Since the local coupling considered here is orbital diagonal in the original orbital basis, the resulting self-energy is diagonal in the same basis,
\begin{equation}
\Sigma(\omega)
=
\begin{pmatrix}
\Sigma_{11}(\omega) & 0 \\
0 & \Sigma_{22}(\omega)
\end{pmatrix}.
\end{equation}
This self-energy is inserted back into the lattice Dyson equation, and the process is repeated until convergence.

For the electron-phonon calculation, we use a local Holstein-type coupling to a local optical phonon mode of frequency \(\omega_0=1\).
The coupling strength is set to \(g=0.3\).
Within the Migdal approximation, the diagonal components of the self-energy are given by
\begin{equation}
\Sigma_{{\rm ph},aa}(t)
=
i g^2
D(t)G_{{\rm loc},aa}(t),
\qquad a=1,2 ,
\end{equation}
where \(D\) is the bare phonon propagator.
The lesser component is
\begin{equation}
\Sigma_{{\rm ph},aa}^{<}(t)
=
i g^2
D^{<}(t)G_{{\rm loc},aa}^{<}(t),
\label{eq:SM_ph_lesser}
\end{equation}
and the retarded component is obtained by the Langreth rule,
\begin{equation}
\begin{aligned}
\Sigma_{{\rm ph},aa}^{R}(t)
= i g^2 \Big[
& D^{R}(t)G_{{\rm loc},aa}^{<}(t)
  + D^{<}(t)G_{{\rm loc},aa}^{R}(t)  \\
& + D^{R}(t)G_{{\rm loc},aa}^{R}(t)
\Big].
\end{aligned}
\label{eq:SM_ph_retarded}
\end{equation}
These expressions are evaluated self-consistently using the dressed local Green's function.

For the local interaction, we use a boldified second-order perturbation theory with \(U=0.8\)~\cite{Aoki2014}.
The second-order self-energy is evaluated from the dressed local Green's function.
In terms of greater and lesser components, we use
\begin{equation}
\Sigma_{U,aa}^{\gtrless}(t)
=
U^2
\left[
G_{{\rm loc},aa}^{\gtrless}(t)
\right]^2
G_{{\rm loc},aa}^{\lessgtr}(-t),
\label{eq:SM_U_greater_lesser}
\end{equation}
and the retarded component is reconstructed as
\begin{equation}
\Sigma_{U,aa}^{R}(t)
=
\theta(t)
\left[
\Sigma_{U,a}^{>}(t)-\Sigma_{U,a}^{<}(t)
\right].
\label{eq:SM_U_retarded}
\end{equation}
This approximation captures the leading local interaction-induced spectral renormalization and damping effects, while remaining numerically affordable in the frequency-dependent steady-state calculation.

\begin{figure}[!tbp]
    \centering
    \includegraphics[width=\columnwidth]{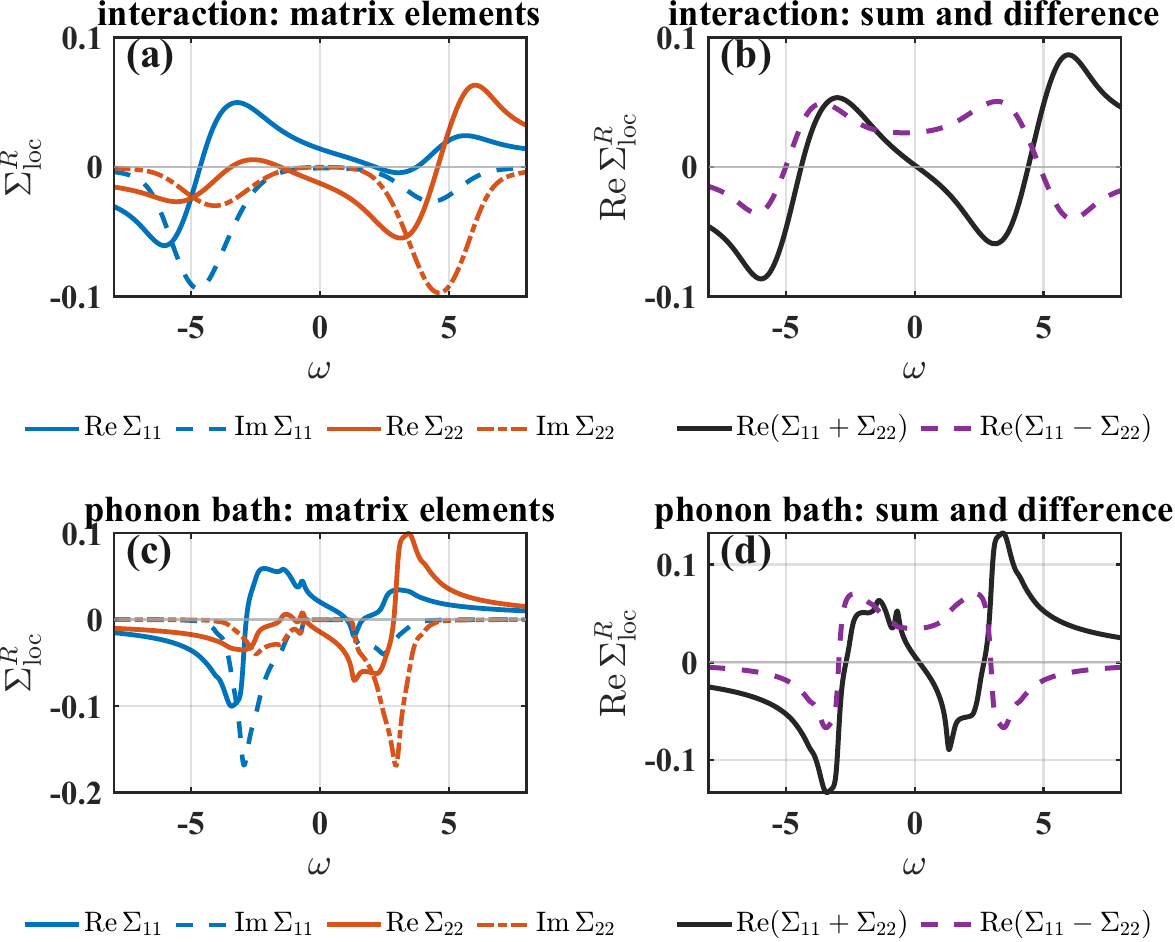}
\caption{
Local retarded self-energies for the interaction and phonon-bath cases.
(a,c) Real and imaginary parts of the diagonal matrix elements $\Sigma^R_{11,\mathrm{loc}}$ and $\Sigma^R_{22,\mathrm{loc}}$.
(b,d) Sum and difference of their real parts, $\mathrm{Re}(\Sigma^R_{11,\mathrm{loc}}+\Sigma^R_{22,\mathrm{loc}})$ and $\mathrm{Re}(\Sigma^R_{11,\mathrm{loc}}-\Sigma^R_{22,\mathrm{loc}})$.
}

\label{fig:SM_self_energy}
\end{figure}

Figure~\ref{fig:SM_self_energy}(a) shows the real and imaginary parts of the two diagonal components of the interaction self-energy for the parameters used in the representative case of the main text: a Lorentzian fermionic bath with peak amplitude \(\Gamma=0.1\) and half-width \(W=4\), at \(T=1/30\) and \(\mu=0.3\), but here we set \(E=0\).
The corresponding sum and difference are shown in Fig.~\ref{fig:SM_self_energy}(b).
Figure~\ref{fig:SM_self_energy}(c) shows the analogous quantities for the electron-phonon self-energy, and Fig.~\ref{fig:SM_self_energy}(d) shows their sum and difference.
The low-energy behavior is similar in the two cases.
The real parts are sizable, while the imaginary parts remain relatively small in the low-energy window relevant for the Hall response.
Thus, for these parameters, the low-energy broadening is still dominated by the fermionic bath damping, whose peak amplitude is \(\Gamma=0.1\), whereas the interaction and electron-phonon self-energies mainly affect the response through their real parts. Since the results shown here are obtained at \(E=0\), they should be viewed as equilibrium reference self-energies for the nonequilibrium calculations. A finite electric field can in principle generate additional low-energy scattering in the driven steady state.
For the phonon self-energy, the imaginary part is strongly suppressed in the window around the chemical potential bounded by the one-phonon threshold.
For the interaction self-energy, the imaginary part is small but not strictly zero.

The real part of the diagonal self-energy can be decomposed into a symmetric and an antisymmetric part in orbital space,
\begin{equation}
\Sigma_0(\omega)
=
\frac{\Sigma_{11}(\omega)+\Sigma_{22}(\omega)}{2},
\qquad
\Sigma_z(\omega)
=
\frac{\Sigma_{11}(\omega)-\Sigma_{22}(\omega)}{2}.
\end{equation}
The symmetric part \(\Sigma_0\) shifts both orbitals equally and therefore mainly produces an overall energy shift.
It does not modify the Berry curvature.
By contrast, the antisymmetric part \(\Sigma_z\) acts as an additional \(\sigma_z\) term.
For the model used in the main text 
\begin{equation}
d_z(\mathbf{k})
=
(\cos k_y-D_y),
\end{equation}
so the real part of \(\Sigma_z\) effectively renormalizes \(D_y\).
At low energy, this can be approximated by
\begin{equation}
D_y^{\rm eff}
=
D_y
-
{\rm Re}\,\Sigma_z^{R}(\omega=0).
\label{eq:SM_Dy_eff}
\end{equation}
A positive value of \({\rm Re}\,\Sigma_{11}^{R}(0)-{\rm Re}\,\Sigma_{22}^{R}(0)\) therefore reduces the effective \(D_y\), opens a small gap near the nodal points, and modifies the Berry-curvature hot spots. We have demonstrated this point in the main text.

\begin{figure}[!tbp]
    \centering
    \includegraphics[width=\columnwidth]{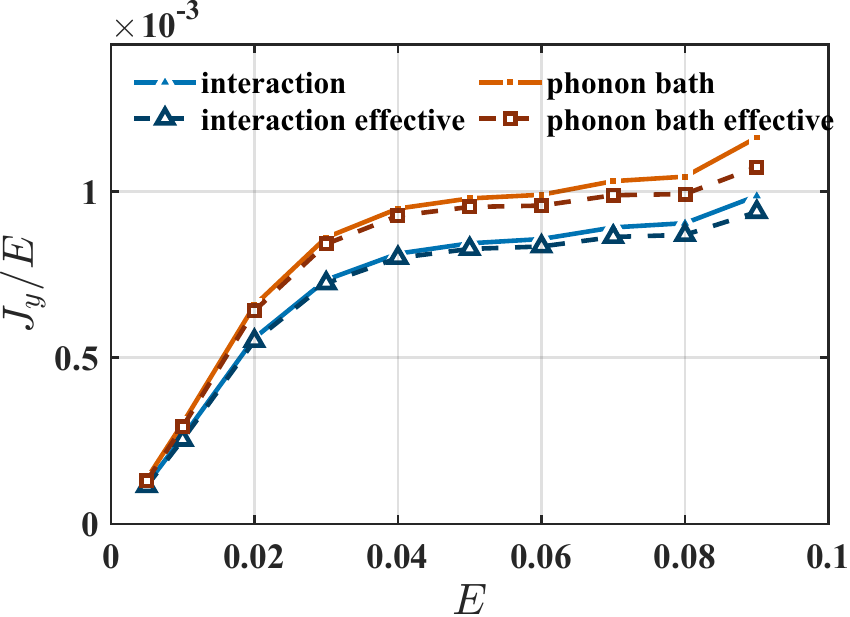}
\caption{
Current response $J_y/E$ as a function of electric field $E$ for the model with electron-electron interactions and phonon-bath, respectively. 
Solid lines show the full calculation, while dashed lines show the corresponding effective descriptions with modified $D_y$.
}
\label{fig:SM_effective_Dy}
\end{figure}

To verify this interpretation further, we compare the full self-consistent calculation with an effective noninteracting calculation in which \(D_y\) is replaced by \(D_y^{\rm eff}\).
The value of \(D_y^{\rm eff}\) is extracted from the zero-field self-energy using Eq.~\eqref{eq:SM_Dy_eff}.
The resulting Hall responses are shown in Fig.~\ref{fig:SM_effective_Dy}.
At relatively small fields, the effective-\(D_y\) calculation reproduces the full electron-phonon and interaction results well.
This confirms that the enhancement of the Hall response mainly originates from the real-part self-energy correction to the Berry-curvature distribution.
At larger fields, deviations appear because the self-energy itself is modified by the driven nonequilibrium distribution and by the self-consistency of the full calculation.
Nevertheless, the overall agreement demonstrates that the dominant effect of both electron-phonon coupling and local interactions in the present regime is the self-energy-induced modification of the low-energy band geometry, rather than a simple change of the scattering rate.

\section{Application to a $C_4\mathcal{T}$-symmetric model}

To test the generality of our conclusions and analysis beyond the
specific model studied in the main text, we apply the NEGF formulation
with the constant-$\Gamma$ reservoir model to the $C_4\mathcal{T}$-symmetric model introduced in
Ref.~\cite{Sur2024FNE},
\begin{align}
H(\mathbf{k})={}&
t_1\cos\frac{k_x}{2}\cos\frac{k_y}{2}\,\sigma_x \notag\\
&+t_2\sin\frac{k_x}{2}\sin\frac{k_y}{2}\,\sigma_y \notag\\
&+t_3\left(\cos k_x+\cos k_y-\Delta\right)\sigma_z \notag\\
&+t_0\left(\cos k_x+\cos k_y\right)\sigma_0 .
\label{eq:SM_C4T_model}
\end{align}
Taking $t_3=1$ as the energy unit, we use the parameters,
\begin{equation}
\begin{aligned}
\Delta&=1.98, &
\frac{t_1}{t_3}&=0.05, &
\frac{t_2}{t_3}&=0.5,\\
\frac{t_0}{t_3}&=0.1, &
\mu&=0.05.&&
\end{aligned}
\label{eq:SM_C4T_parameters}
\end{equation}
Because of $1/2$ factors in
Eq.~\eqref{eq:SM_C4T_model}, the Hamiltonian has a $4\pi$ period in
each momentum direction. We therefore perform the momentum-space
calculation over $k_x,k_y\in[-2\pi,2\pi]$. The reservoir damping
parameter used in our NEGF calculation is $\Gamma=0.1$.

The $C_4\mathcal{T}$ symmetry implies
$H^*(k_y,-k_x)=H(k_x,k_y)$ and forces the Berry-curvature dipole to
vanish. The leading perturbative Hall response is therefore governed
by the Berry-curvature quadrupole, which means $J_y/E\propto E^2$ in the weak-field limit.

\begin{figure}[!tbp]
    \centering
    \includegraphics[width=\columnwidth]{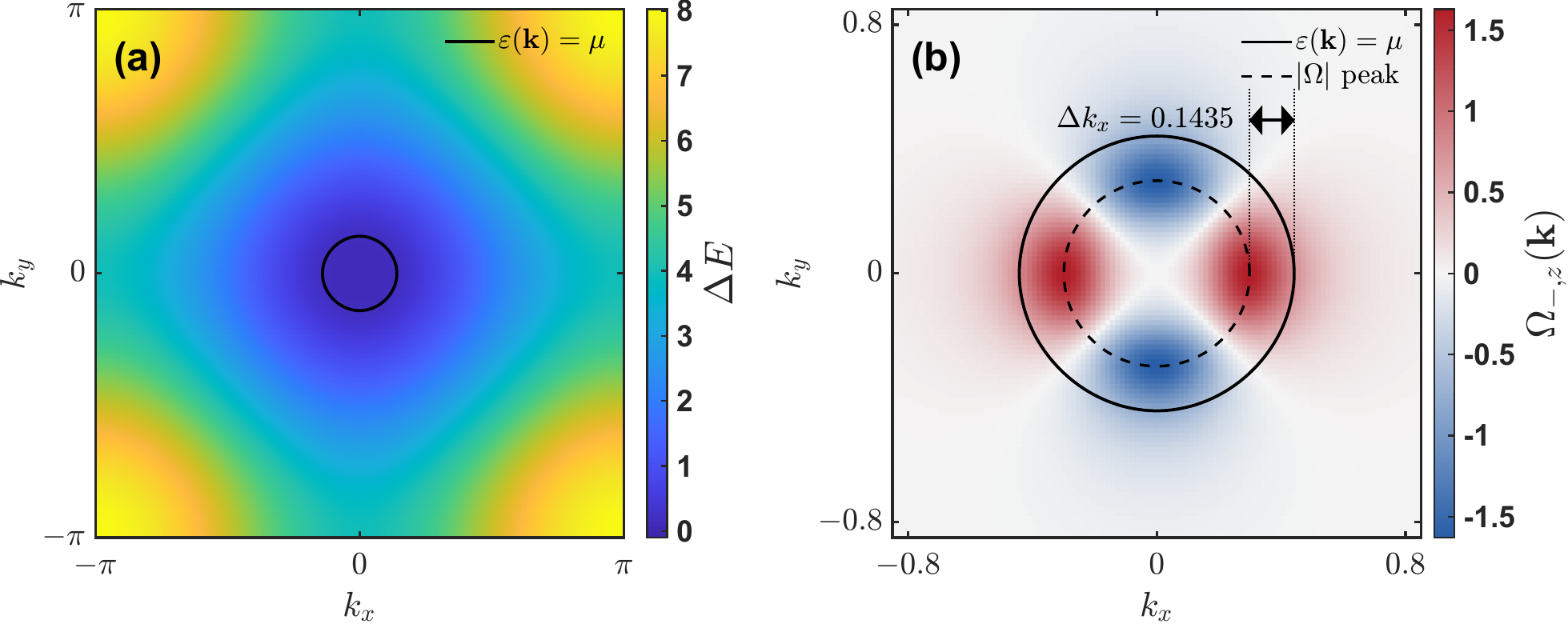}
\caption{
Band geometry of the $C_4\mathcal{T}$-symmetric model in 
momentum space. 
(a) Direct band gap $\Delta E(\mathbf{k})$.
(b) Berry curvature
$\Omega_{-,z}(\mathbf{k})$. The solid black curve denotes the
equilibrium Fermi contour, while the dashed curve in panel (b) marks
the peaks of the Berry-curvature hot spots. Their separation along the
field direction is $\Delta k_x=0.1435$.
}
\label{fig:SM_C4T_model}
\end{figure}

Figure~\ref{fig:SM_C4T_model} shows the band gap and Berry
curvature together with the equilibrium Fermi contour. For a field
applied along $x$, the closest Berry-curvature hot spot lies a distance
$\Delta k_x=0.1435$ from the Fermi-contour edge. The estimate
for the crossover field is therefore $E=\Gamma\Delta k_x\simeq0.0143$.

\begin{figure}[!tbp]
    \centering
    \includegraphics[width=0.9\columnwidth]{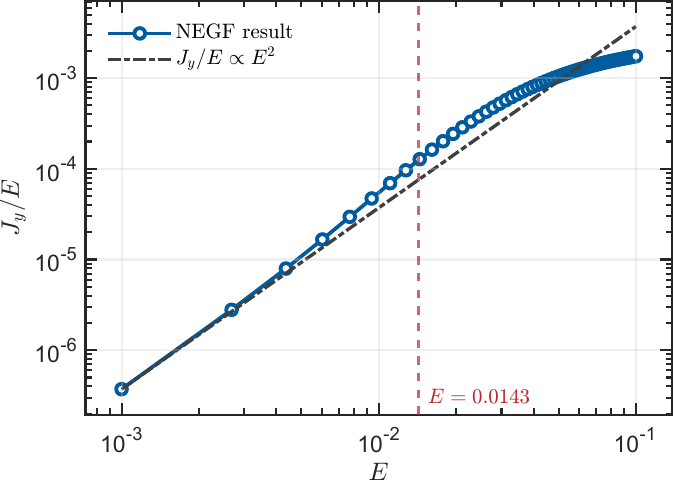}
\caption{
Field-normalized Hall current $J_y/E$ of the
$C_4\mathcal{T}$-symmetric model, calculated with $\Gamma=0.1$.
The dashed line indicates the perturbative Berry-curvature-quadrupole
scaling $J_y/E\propto E^2$, and the vertical line marks the estimated
crossover field $E=\Gamma\Delta k_x\simeq0.0143$.
}
\label{fig:SM_C4T_Hallresponse}
\end{figure}

As shown in Fig.~\ref{fig:SM_C4T_Hallresponse}, the NEGF result follows
the expected $J_y/E\propto E^2$ scaling at weak fields. It begins to
deviate clearly from this perturbative behavior around the estimated
crossover field, where the driven Fermi-contour edge reaches the
Berry-curvature hot spot. This crossover is qualitatively similar to
that reported in Ref.~\cite{Sur2024FNE}, where the quadrupolar
weak-field scaling likewise breaks down as the field is increased.
This example demonstrates the generality of our analysis based on the
field-driven redistribution relative to Berry-curvature hot spots, as
well as the broader applicability of our conclusion concerning the
breakdown of perturbative scaling. These features persist when
symmetry eliminates the Berry-curvature dipole and the leading
weak-field response is instead controlled by the Berry-curvature
quadrupole.

\section{Numerical convergence tests}

\begin{figure}[!tbp]
    \centering
    \includegraphics[width=\columnwidth]{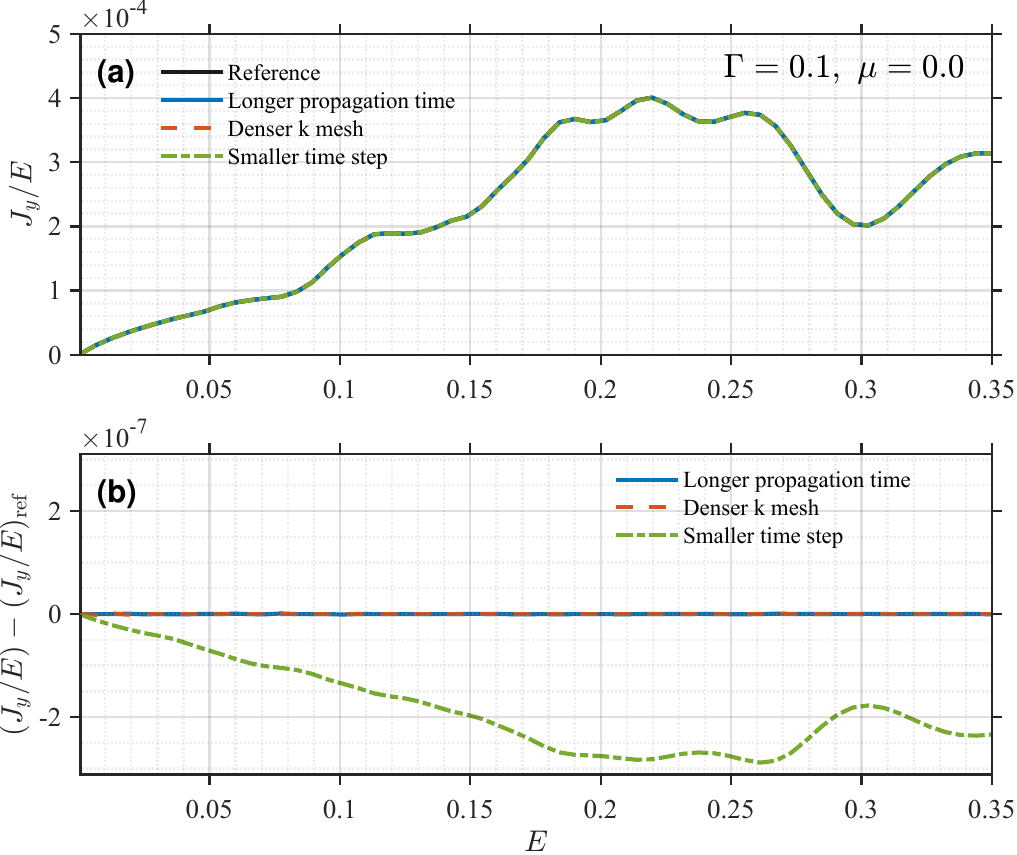}
    \caption{
    Numerical convergence of the field-normalized Hall current
    $J_y/E$ for $\mu=0$ and $\Gamma=0.1$ in the constant-$\Gamma$
    reservoir model. (a) Reference result compared with calculations
    using twice the maximum propagation time, twice the momentum-grid
    resolution, and half the time step. The four curves are
    indistinguishable on the scale of the plot. (b) Difference between
    each test result and the reference calculation. The
    largest absolute deviation, produced by halving the time step, is
    of order $10^{-7}$; the other two deviations are approximately two
    orders of magnitude smaller.
    }
    \label{fig:SM_hallresponse_convergence}
\end{figure}

We verify the numerical convergence of representative main-text
results for both reservoir implementations. For the constant-$\Gamma$
reservoir model, we use the $\mu=0$ and $\Gamma=0.1$ field-normalized
Hall current in Fig.~1(b) of the main text as a representative example. Starting from the reference calculation, we independently double the maximum propagation time, halve the time step, and double the momentum-grid resolution. As shown in Fig.~\ref{fig:SM_hallresponse_convergence}, all three
modified calculations overlap with the reference curve. The numerical
deviations are negligible compared with the $10^{-4}$ scale of the
Hall response, demonstrating that the main-text results obtained with
the constant-$\Gamma$ reservoir model are well converged.

\begin{figure}[!tbp]
    \centering
    \includegraphics[width=\columnwidth]{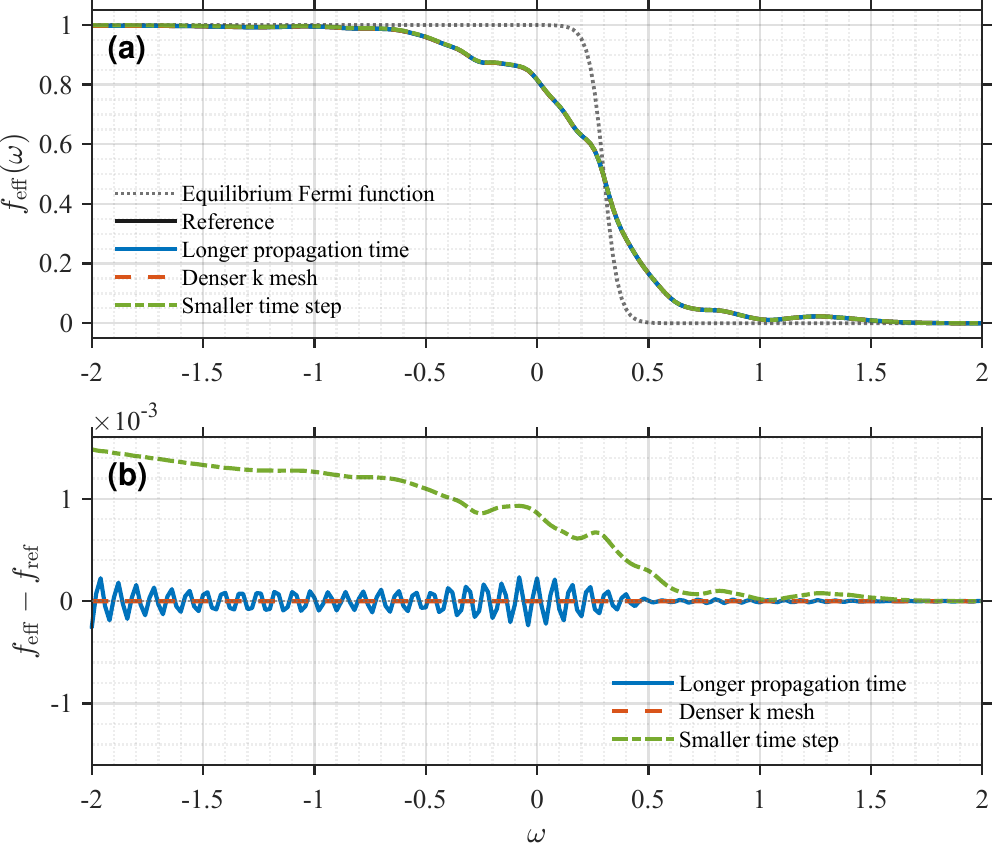}
    \caption{
    Numerical convergence of the effective distribution function
    $f_{\mathrm{eff}}(\omega)$ at $E=0.15$ for the frequency-dependent
    Lorentzian fermionic reservoir. (a) Reference result compared with
    calculations using twice the maximum propagation time, twice the
    momentum-grid resolution, and half the time step. The equilibrium
    Fermi function is shown for reference. (b) Difference between each
    convergence-test result and the reference calculation. The
    deviations remain of order $10^{-3}$ or smaller.
    }
    \label{fig:SM_distribution_convergence}
\end{figure}

We perform the same tests for the frequency-dependent Lorentzian
fermionic reservoir used in Fig.~3 of the main text. We focus on the
effective distribution function at $E=0.15$, where its nonequilibrium
structures are most pronounced, and independently double the maximum
propagation time and momentum-grid resolution and halve the time step. Figure~\ref{fig:SM_distribution_convergence} shows that the
effective distribution functions obtained with all the settings overlap 
on the scale of the plot. Therefore,
the nonequilibrium structures remain unchanged within the numerical
accuracy. Together, Figs.~\ref{fig:SM_hallresponse_convergence} and
\ref{fig:SM_distribution_convergence} demonstrate that the main-text
results are converged for both the constant-$\Gamma$
reservoir and the frequency-dependent reservoir.

\end{document}